\documentclass[preprint2]{aastex63}
\usepackage{amsmath}
\usepackage{CJK}
\usepackage{listings}
\usepackage{xcolor}
\usepackage{fontawesome5}
\usepackage{ulem}

\definecolor{codegreen}{rgb}{0,0.6,0}
\definecolor{codegray}{rgb}{0.5,0.5,0.5}
\definecolor{codepurple}{rgb}{0.58,0,0.82}
\definecolor{backcolour}{rgb}{0.95,0.95,0.92}

\lstdefinestyle{mystyle}{
  backgroundcolor=\color{backcolour},   commentstyle=\color{codegreen},
  keywordstyle=\color{magenta},
  numberstyle=\tiny\color{codegray},
  stringstyle=\color{codepurple},
  basicstyle=\ttfamily\footnotesize,
  breakatwhitespace=false,         
  breaklines=true,                 
  captionpos=b,                    
  keepspaces=true,                 
  numbers=left,                    
  numbersep=5pt,                  
  showspaces=false,                
  showstringspaces=false,
  showtabs=false,                  
  tabsize=2
}

\newcommand{\AB}[2]{$\mbox{[#1/#2]}$}

\newcommand{\teff}{$T_{\rm eff}$}
\newcommand{\logg}{$\log g$}
\newcommand{\vt}{$v_{\mathrm{mic}}$}
\newcommand{\kms}{km\,s$^{-1}$}

\newcommand{\lotus}{\texttt{LOTUS}}

\DeclareRobustCommand{\ion}[2]{\textup{#1\,\textsc{\lowercase{#2}}}}

\newcommand\fei{\ion{Fe}{i}}
\newcommand\feii{\ion{Fe}{ii}}
\newcommand{\feh}{\AB{Fe}{H}}

\received{August 23, 2022}
\revised{January 09, 2023}
\accepted{January 27, 2023}
\submitjournal{AJ}

\shorttitle{NLTE Stellar Parameters with \lotus}
\shortauthors{Li et al.}
\graphicspath{{./}{figures/}}
\begin{document}
\begin{CJK*}{UTF8}{gbsn}
\title{\texttt{LOTUS}: A (non-)LTE Optimization Tool for Uniform derivation of \\Stellar atmospheric parameters}

\correspondingauthor{Yangyang Li}
\email{yangyangli@ufl.edu}

\author[0000-0002-9953-7929]{Yangyang Li(李扬洋)}
\affiliation{Department of Astronomy, University of Florida, Bryant Space Science Center, Gainesville, FL 32611, USA}

\author[0000-0002-8504-8470]{Rana Ezzeddine}
\affiliation{Department of Astronomy, University of Florida, Bryant Space Science Center, Gainesville, FL 32611, USA}
\affiliation{Joint Institute for Nuclear Astrophysics - Center for Evolution of the Elements, USA}




\begin{abstract}

%It has been shown that Non-Local Thermodynamic Equilibrium (NLTE) stellar spectral models are needed to determine precise stellar fundamental atmospheric parameters and chemical abundances, which constitute the cornerstone of understanding the history and evolution of stars and galaxies. %The classical and most common methods to determine chemical elemental abundances in stars are based on measurements of equivalent widths (EWs) or the computation of synthetic absorption line spectra of the chemical elements in question.
%However, due to the highly expensive computational time of 
Precise fundamental atmospheric stellar parameters and abundance determination of individual elements in stars are important for all stellar population studies. Non-Local Thermodynamic Equilibrium (Non-LTE; hereafter NLTE) models are often important for such high precision, however,  can be computationally complex and expensive, which renders the models less utilized in spectroscopic analyses. 
%Nowadays, NLTE abundance determinations based on pre-computed equivalent widths (EW) or departure coefficient grids have started to become more commonly available.
%, it has been less common to  measure abundances recursively via spectral fitting synthetic spectra. %Even with the help of using pre-computed grids of synthetic spectra, large spectra library is still a heavy burden for data storage.
%Abundance determinations based on EW has been shown to be promising avenue for NLTE chemical abundance determination from high resolution spectra. 
To alleviate the computational burden of such models, we developed a robust 1D, NLTE fundamental atmospheric stellar parameter derivation tool, \texttt{LOTUS}, to determine the effective temperature \teff, surface gravity \logg, metallicity [Fe/H] and microturbulent velocity \vt\ for FGK type stars, from equivalent width (EW) measurements of \fei\ and \feii\ lines. 
%of inferred stellar parameters and iron abundance more accurately. 
We utilize a generalized curve of growth method to take into account the EW dependencies of each \fei\ and \feii\ line on the corresponding atmospheric stellar parameters. A global differential evolution optimization algorithm is then used to derive the fundamental parameters. Additionally, \texttt{LOTUS} can determine precise uncertainties for each stellar parameter using a Markov Chain Monte Carlo (MCMC) algorithm. 
We test and apply \texttt{LOTUS} on a sample of benchmark stars, as well as stars with available asteroseismic surface gravities from the K2 survey, and metal-poor stars from the $Gaia$-ESO and $R$-process Alliance (RPA) surveys. We find very good agreement between our NLTE-derived parameters in \texttt{LOTUS} to non-spectroscopic values on average within \teff\ $= \pm 30$\,K and \logg\ $= \pm 0.10$\,dex for benchmark stars. 
%The uncertainties of our models are to within \teff$=\pm 70$\,K, \logg$= \pm 0.10$\,dex,  [Fe/H]$= \pm 0.05$\,dex and $\xi_t= \pm 0.07$\kms. 
We provide open access of our code, as well as of the interpolated pre-computed NLTE EW grids available on Github \href{https://github.com/Li-Yangyang/LOTUS}{\faIcon{github}}\footnote{The software is available on GitHub \url{https://github.com/Li-Yangyang/LOTUS} under a MIT License and version 0.1.1 (as the persistent version) is archived in Zenodo \citep{yangyang_li_2023_7502804}.}, and documentation with working examples on Readthedocs \href{https://lotus-nlte.readthedocs.io/en/latest/}{\faIcon{book}}.

\end{abstract}



\section{Introduction} \label{sec:intro}
\end{CJK*}
Precise characterization of stellar spectra is a key ingredient for understanding several fields of modern astrophysics, including the physical and chemical properties and abundances of stars \citep{Asplund2009, Jofre2019}, galactic formation and evolution \citep{Audouze1976, McWilliam1997, Kobayashi2006}, as well as macro and micro physical phenomena near the surface of stars \citep{Miesch2009, Linsky2017}. This accurate level of characterization is also needed for the detection of extra-solar planets when using radial velocity methods \citep{Vanderburg2016}.

With the current inflow of stellar spectra from ongoing and future large observational spectroscopic surveys including,  SDSS-V (\citealt{Kollmeier2017}, R$\sim$2000 and $\sim$22500), LAMOST (\citealt{Liu2020}, R$\sim$7500; \citealt{Cui2012, Zhao2012, Deng2012}, R$\sim$1800), APOGEE (\citealt{Ahumada2020}, R$\sim$22500), RAVE (\citealt{Steinmetz2020}, R$\sim$7500), GALAH (\citealt{DeSilva2015}, R$\sim$28000), as well as the upcoming WEAVE (\citealt{Dalton2016}, R$\sim$5000 and R$\sim$20000), 4MOST (\citealt{deJong2019}, R$\sim$20000) and PLATO (\citealt{Miglio2017}), stellar spectroscopy will be providing $\sim 10^5$ of golden opportunities to study the chemical and dynamical properties of stars in the Galaxy to help understand its build-up history and evolution. 

Atmospheric fundamental stellar parameters, including the effective temperatyre \teff, surface gravity \logg, metallicity [Fe/H] and microturbulent velocity \vt, as well as chemical abundances of stars, are determined from the observed spectra of a given star by fitting theoretical synthetic spectra based on assumptions of geometric structures (1D versus 3D), radiative transfer assumptions (Local Thermodynamic Equilibrium; hereafter LTE, versus Non-Local Thermodynamic Equilibrium; hereafter NLTE).
%and assumed fundamental stellar parameters (\teff, \logg, [Fe/H] and \vt).
%THERE"S NO NEED FOR THIS SENETEceXX or geometric description, commonly used 1D models (MARCS;  \citealt{Gustafsson2008} and  ATLAS9; \citealt{C&K2003}), which assume plane-parallel or spherical geometry for each layer of the atmosphere, and 3D models (STAGGER; \citealt{Magic2013} and CO$^5$BOLD; \citealt{Freytag2012}), which is more realistic than 1D. For stellar parameters, we generally classified them into two categories, one includes stellar atmospheric parameters, such as effective temperature T$_{\mathrm{eff}}$, surface gravity \logg, metallicity [Fe/H]; the other accounts for line broadening , such as micro-turbulence velocity $\xi_t$. Noticeably, $\xi_t$ is an non-physical parameter which is needed to adjust for the shape of line in 1D models. Hence, this parameter doesn't have to be a key parameter when comparing with different methods, which is not required in 3D full hydrodynamic simulations \citep{Jofre2019}. 
Two classical methods are commonly used to derive stellar atmospheric parameters from stellar spectra: either 1) by iteratively fitting the observed spectra to synthetic spectral models until a best-fit match is met at the corresponding stellar parameters, or 2) by measuring chemical abundances determined from \fei\ and \feii\ lines from equivalent widths (EW) measurements, and employing optimization of excitation and ionization equilibrium by changing the stellar parameters iteratively until trends with excitation potential energies ($\chi$) and reduced equivalent widths ($\log (EW/\lambda)$) of the lines are minimized. While the former method of spectral synthesis might outperform the latter in crowded spectral regions or spectra dominated by strong lines ($\log(EW/\lambda)>-4.5$), the EW method generally requires less calculation time and resources than synthetic spectra, making it a simpler and more widely applied tool to measure atmospheric stellar parameters and chemical abundances.
%In theory, if our models consider all possible physical scenarios and include all possible noise from instruments and Earth's atmosphere, there should be a single theoretical model that matches with the observed spectrum of a given star. 

Classical radiative transfer models for deriving stellar parameters and chemical abundances from observed spectra most often LTE. Under LTE, the gas and chemical particles throughout the stellar atmosphere satisfy the Saha-Boltzmann excitation and ionization balance equations. Several stellar atmospheric parameter tools implementing classical 1D and LTE approximations for spectroscopic analyses already exist and are widely used (e.g., MATISSE \citep{Recio-Blanco2006}, MyGIsFOS \citep{Sbordone2014}, the APOGEE pipeline ASCAP \citep{Garcia2016}, GALA \citep{Mucciarelli2013}, DOOp \citep{CG2014}, ARES \citep{Sousa2015}, StePar \citep{Tabernero2019}, FASMA \citep{Tsantaki2021} and iSpec \citep{BC2014, BC2019}.

The LTE assumption utilized by all of these tools is, however, not physically motivated when applied to evolved giants and/or metal-poor stars \citep{Lind2012, Ezzeddine2017, Mashonkina2017, Amarsi2020}. Inelastic collisional interactions between atoms and electrons or neutral hydrogen in the atmospheres of cool stars usually drive the local conditions in these atmospheres towards LTE. In metal-poor and giant stars, limited electron donors from metals 
are not able to induce enough collisions to maintain collisional equilibrium in the line-forming regions \citep{Hubeny2014}, which drive line formation away from LTE \citep{Mihalas1973, Bergemann2012}. Quantified deviations in abundances and stellar atmospheric parameters from LTE are commonly known as NLTE effects \citep{Asplund2005}, which are mainly driven by deviations from statistical equilibrium or kinematic equilibrium. 

A small number of radiative transfer codes have started taking NLTE effects into account for the determination of chemical abundances, including \texttt{PySME} \citep{pysme}, which is based on the spectroscopic analysis code Spectroscopy Made Easy (SME; \citealt{sme}). \citet{Kovalev2019} has implemented NLTE departure coefficients in the synthesis tools, allowing NLTE spectral synthesis for certain corresponding absorption lines. More recently, a NLTE version of Turbospectrum \citep{TSplez2012} has been released \citep{gerber2022}, which also has NLTE departure coefficients incorporated in their 1D, LTE spectral synthesis models. However, both codes allow NLTE corrections for only some elements via spectral synthesis, and stellar parameters still need to be derived independently or iteratively via full spectral fitting, which can be prone to degeneracies. Therefore, a tool that can automatically derive stellar parameters incorporating NLTE models, taking into account stellar parameter dependencies, is highly needed which is the main motivation of this work and paper.

%a motivation why our code is needed that it is the first tool analysing spectra via EW method based on NLTE calculation.

In this work, we present a fast, automatic and robust spectroscopic analysis tool, \texttt{LOTUS}, which stands for a ``(non-)LTE Optimization Tool for Uniform derivation of Stellar atmospheric parameters''. \texttt{LOTUS} allows the derivation of \teff, \logg, [Fe/H] and \vt\ based on EW measurements input of \fei\ and \feii\ lines from stellar spectra. Either or both LTE and NLTE modes can be chosen for the benefit of comparisons. As compared to traditional full NLTE calculations for each model, \texttt{LOTUS} can significantly shorten the time of the determination of stellar parameters using the NLTE assumption from several hours or even days \citep{Hauschildt1997} to $\sim 15-30$ mins for each star. A Monte-Carlo Markov Chain (MCMC) analysis is also implemented to precisely estimate the uncertainties of the derived parameters. We test and apply \texttt{LOTUS} on several benchmark stars, and stellar surveys with available non-spectroscopic atmospheric parameters (from asteroseismology or interferometry)for comparison and validation of our results. 
%The comparison results show high accuracy and prove that NLTE outperforms in accuracy than LTE.

The rest of the paper is organized as follows: In Section \ref{sec:method}, we present a detailed description of \texttt{LOTUS}, and describe the input models for our NLTE calculations, as well as a description of the different modules of the code. %We also demonstrate the interpolation technique adopted in the code to optimize the fits between the measured EW and the theoretical pre-computed EW. 
In Section \ref{sec:test}, we test our code and apply it to derive and compare the derived parameters of benchmark stars, and stars with  non-spectroscopic derived parameters. In Section\,\ref{sec:application}, we apply the code to a large sample of metal-poor stars and discuss the NLTE effects obtained via \lotus. In Section \ref{sec:caveats}, we present a discussion on the caveats and limitations of the code, and finally, in Section \ref{sec:summary} we present a summary of our results and conclude.

\section{General description} \label{sec:method}

%\re{Need to mention what the observation sample is, why did we choose it, and how/where have the EW been adopted from in the analysis}

%EWs are measured from either fitting a normalized spectral profile with Gaussian profile for weak lines and Vogit profile for strong lines, or integrating directly. 

\texttt{LOTUS} is designed to derive the fundamental atmospheric stellar parameters \teff, \logg, [Fe/H] and \vt of FGK type stars, by implementing observed measurements of EW for \fei\ and \feii\ lines as input. An example of a user input EW linelist is shown in Listing \ref{lst:example_input} below. The transition wavelength (\texttt{obs\_wavelength} in {\AA}), element and ionization stage (\fei\ or \feii), measured EW (\texttt{obs\_ew} in m{\AA}) and excitation potential energy EP (\texttt{obs\_ep} in eV) are required. 

\begin{lstlisting}[language=bash,label={lst:example_input}, caption={Example of an input file format of observed EW measurements for \fei\ and \feii\ lines provided by users for \texttt{LOTUS}. \texttt{obs\_wavelength} are in {\AA}, \texttt{obs\_ew} in m{\AA}, and \texttt{obs\_ep} in eV.}]
obs_wavelength,element,obs_ew,obs_ep
4787.8266,FeI,44.2,3.00
4788.7566,FeI,65.5,3.24
4789.6508,FeI,83.3,3.55
6456.3796,FeII,59.2,3.9
...
\end{lstlisting}

\texttt{LOTUS} has three general modules of functionality.
%We describe each of these modalities in detail in Section\,\ref{ssec:modules}}. 
In a general overview,  (i) \fei\ and \feii\ abundances are derived by interpolating a generalized curve of growth (GCOG) for each line in a pre-computed grid of theoretical EW in both LTE and NLTE, following \citet{Boeche2016}. (ii) the stellar parameters are then derived by minimizing the slopes for excitation and ionization equilibrium, iteratively using a global minimization module, and finally, (iii) uncertainties of the derived atmospheric parameters are estimated utilizing a Markov Chain Monte Carlo (MCMC) algorithm. 
%In the next section, we describe the input models 
We describe each of these modules in detail in  sub-section\,\ref{ssec:modules} below, after listing the input and radiative transfer models used in Section\,\ref{ssec:input_models}.

%when there exists no variation between observed EWs and chemical abundances (excitation balance) and no variation abundances between different ions of same element (ionization balance). The physics of both equilibrium condition based on the curve of growth, which describe the relationship between EWs and column density of species, and Saha and Boltzmann equations.

\begin{figure}[ht!]
\epsscale{1.15}
\plotone{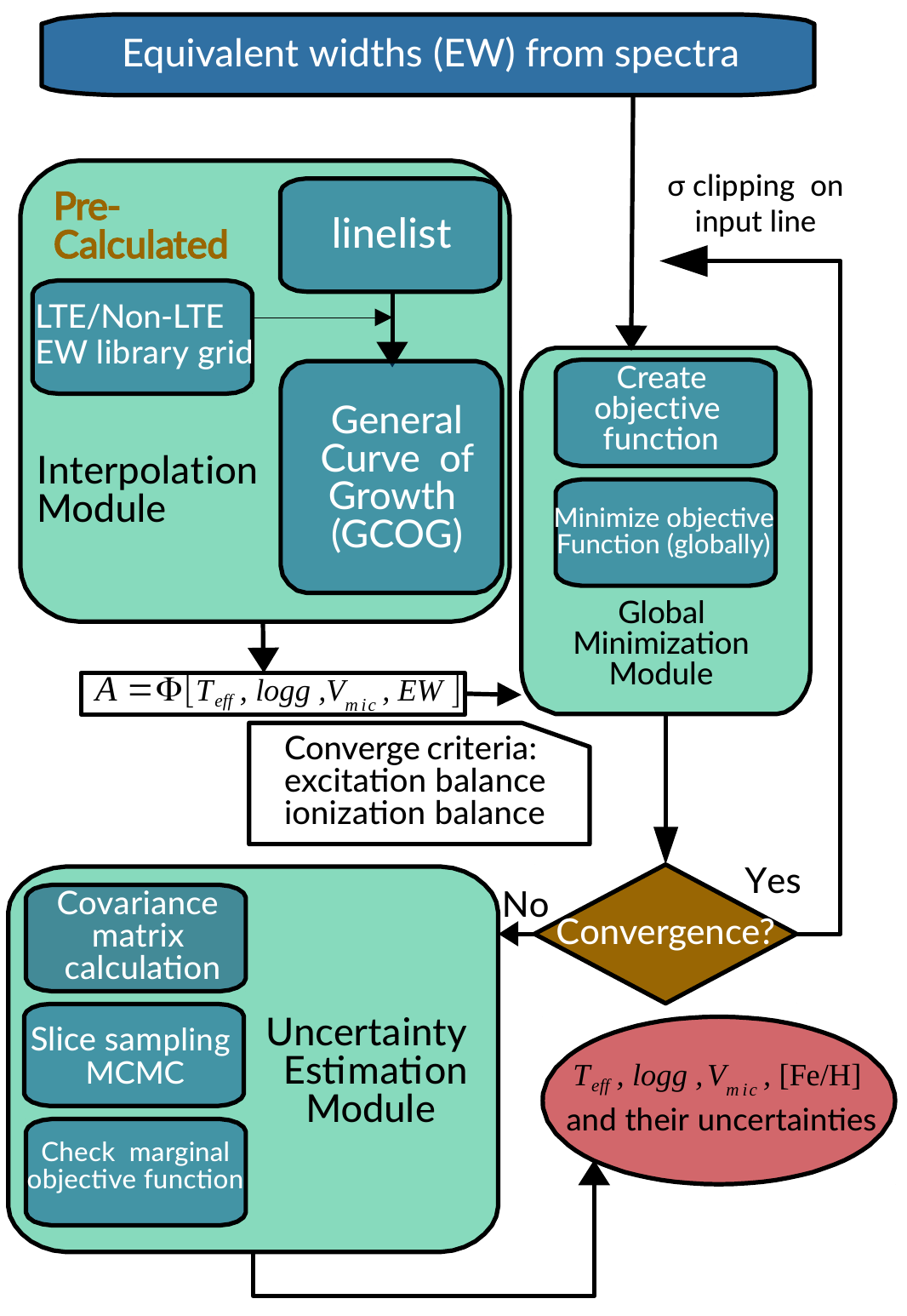}
\caption{
A flow diagram describing the working modules of \texttt{LOTUS}. Three main modules are used to derive the stellar parameters and their uncertainties:  (i) an interpolation Module, (ii) a Global Minimization Module, and (iii) an Uncertainty Estimation Module. Input includes EW measurements of \fei\ and \feii\ of the absorption lines measured in the spectra. Users can define the number of $\sigma$ clipping required to remove outliers (see Section\,\ref{sec:glob_optim} for details). Additionally, a halt condition for the value of the smallest allowable minimization function threshold can also be defined by users (see Section\,\ref{sec:glob_optim}  for details).
%Measured with wavelength and excitation potential are needed as input data. It is worth noting that we calculated interpolation module ahead in order to save time for late procedures. XX these don't need to be in the caption XX
\label{fig:flowchart}}
\end{figure}

%\begin{enumerate}
%    \item A pre-computed grid of LTE and NLTE theoretical EW implementing 1D, MARCS \citep{gustafsson1975,Gustafsson2008} atmospheric models and covering typical FGK range of stellar parameters. The EW are computed using the radiative transfer code \texttt{MULTI2.3} \citep{Carlsson1986} under assumption of both LTE and NLTE. More details are presented in Section \ref{sect:multi}.
%    \item Model atom for Fe, including information of energy levels, radiative and collisonal transition rates for \fei and \feii, which are described in \citet{Ezzeddine2016,Ezzeddine2020}.
%    \item A pre-vetted linelist of \fei\ and \feii\ lines, including up-to-date atomic data implemented from the $Gaia$-ESO survey and XX where else? RPA?XX covering lines expected to be observed in a wide range of \feh\ from XX fill XX.
%    \item A robust interpolation module implementing XX WHAT? XX to compute theoretical EWs for any combination of stellar parameters other than those on the nodes of grid (see Section \ref{sect:ew_interp}).
%    \item Global optimization implementing differential evolution algorithm \citep{Storn1997} (see Section \ref{sect:optimization})
 %   \item Uncertainty estimator module implementing a co-variance matrix and Markov Chain Monte-Carlo (MCMC) method (see Section \ref{sssec:uncert_estimation}).
%\end{enumerate}
\subsection{Input Models}\label{ssec:input_models}
\subsubsection{Stellar Atmosphere Models}\label{sect:atmos_models}
\texttt{LOTUS} incorporates 1D, LTE MARCS stellar atmospheric models \citep{gustafsson1975,Gustaf2008}, covering a wide range of stellar parameters typical for FGK stars. 
%The effective temperatures, T$_{\mathrm{eff}}$, in total range from 2500 to 8000\,K. The grid of models are in steps of 100\,K from 2500\,K to 4000\,K and 250\,K between 4000\,K and 8000\,K while range the logarithm surface  gravity \logg from -0.5 to 5.5 in the steps of 0.5. The range of adopted metallicity is from -5.0 to 1.0 but in variable steps. The option for different micro-turbulence velocity $\xi _{t}$ is also supported. 

%\begin{figure}[ht!]
%\hspace*{-0.8cm}
%\includegraphics[scale=0.5]{interp_comparison.pdf}
%\caption{\textbf{An example interpolation of atmosphere model done MARCS interpolation subroutine, which shows high consistency with its corresponding original MARCS model.}\label{fig:marcs_comp}}
%\end{figure}

Spherical atmospheric models were used for \logg$\,<3.5$, otherwise, plane-parallel models were adopted. 
The grid of MARCS model atmosphere available online\footnote{https://marcs.astro.uu.se/} offers reasonable coverage for the stellar parameters, however they exhibit wide gaps in effective temperature (steps of $\sim 250$\,K), surface gravity (steps of $\sim 0.5$\,cgs), and metallicity (steps of $\sim 0.25 - 0.50$\,dex) to optimally explore the parameter space.
We utilize the MARCS interpolation subroutine \texttt{iterpol\_marcs.f}\footnote{http://marcs.astro.uu.se/software.php} written by Thomas Masseron, to produce a higher resolution grid, with our final parameter grid ranging 
from 4000\,K to 6850\,K for \teff\ (steps of 50\,K), \logg\ from 0.0 to 5.0 (steps of 0.1), [Fe/H] from $-3.5$ to $+0.5$ (steps of 0.5) and \vt\ from 0.5 to 3.0 \kms\ (steps of 0.5\,\kms). 
%\textbf{We show an example of an interpolated atmospheric model using \texttt{iterpol\_marcs.f} in Figure \ref{fig:marcs_comp}, compared to its original MARCS model at \teff=4500\,K, \logg=1.0, \vt=2.0\,\kms and \feh=$-$2.5. Comparison of the temperature profile in the atmosphere as a function of optical depth at 5000\,{\AA} shows excellent agreement which demonstrates that the interpolation works very well.}

%We use 56 depth points in the final interpolated stellar atmosphere models.

\subsubsection{Iron Model Atom}\label{sect:model_atom}
Computing theoretical NLTE \fei\ and \feii\ EWs requires the input of a comprehensive iron model with up-to-date radiative and collisional atomic data. For our EW grid calculations, we adopt a well-tested \fei/\feii\ model atom containing 846 \fei\ and 1027 \feii\ lines \citet{Ezzeddine2016,Ezzeddine2017,Ezzeddine2020}. 
%$\log gf$ values were ado[ted and van der Waals collisional broadening parameters are taken from GES line list v5 \citep{Jofre2014, Heiter2015, Ezzeddine2020}. 
The atom includes absorption line transitions spanning the near-UV to the near-IR, extending the range of wavelength from $\sim 1000$ to $10^5$\,{\AA}. The model also carefully considers hydrogen collision and electron collision processes from quantum atomic data, particularly implementing ion-pair production and mutual-neutralization processes from \citet{barklem2018}. Extensive details on the build-up of the atom and the corresponding atomic data are presented in \citet{Ezzeddine2016} and \citet{Ezzeddine2020}.

\subsubsection{NLTE Radiative Transfer computations}\label{sect:multi}
%Since the complexities when solving radiative transfer equation and statistical equilibrium under the assumption of LTE or NLTE, MULTI2.3 is the ideal tool to compute spectral profile with various non-linear atomic processes coupling. 
We utilized the NLTE radiative transfer code \texttt{MULTI2.3} to solve for statistical equilibrium populations and derive the theoretical NLTE EW for the \fei\ and \feii\ lines in our atom.
The code utilizes the Approximate Lambda Iteration (ALI) \citep{Rybicky1991} to iteratively determine the populations using the comprehensive \fei/\feii\ model atom described in Section\,\ref{sect:model_atom}. %For each line, the calculation of EW$_{\mathrm{NLTE}}$ needs 600 frequency points as a maximum ?? xwhat is . The maximum are considered as 80000 and 2000 individually for bound-bound transitions and bound-free transitions. The maximum angle points when integrating along the solid angle are 5. 
\texttt{MULTI2.3} also solves for LTE populations using the classical Saha and Boltzmann equations, which is output as a departure coefficient, $b_i$, where $b_i=n_i(\mathrm{NLTE})/n_i(\mathrm{LTE})$ \citep{Wijbenga1972}, where $n$ is the level population for the corresponding line transition $i$. 

\subsubsection{Linelist Selection}\label{sec:linelist}
The linelist selection is crucial for the accurate derivation of \fei\ and \feii\ chemical abundances and stellar parameters using the EW method, especially for cooler stars (e.g. FGK stars) since they have much denser spectral line regions containing blended lines. Therefore, we choose a comprehensive list of \fei\ and \feii\ from the $Gaia$-ESO line list \citep{Jofre2014, Heiter2015, Heiter2021}, covering the wavelength ranges from 4750 to 6850\,{\AA} and from 8500 to 8950 {\AA}. Additional lines from the R-Process Alliance (RPA) survey (e.g., \citealp{Hansen2018, Sakari18, Ezzeddine2020,rpa4}) have also been added to account for lines common in metal-poor stars, with corresponding atomic data and references in \citep{Roederer_2018}. We combine iron lines from both linelists, removing any duplicates in the process. Our final linelist is presented in Table\,\ref{tab:linelist}. Future \lotus\ releases should be easily able to extend the linelists to bluer and redder lines in the UV and IR, respectively.

\begin{figure*}
\plotone{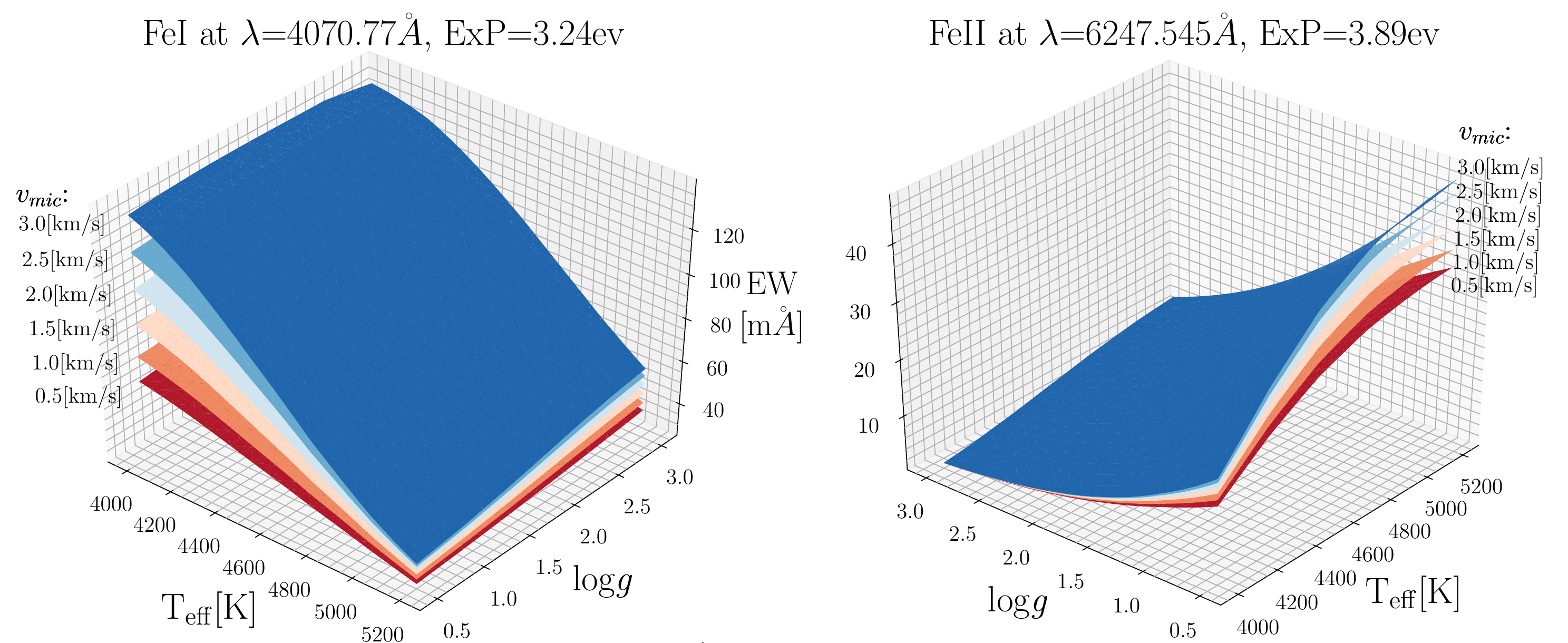}
\caption{Computed NLTE EW dependence as a function of \teff, \logg, \vt\ for the Fe\,I line at 4070.77{\AA} and EP=3.24\,eV (left) and the Fe\,II line at 6247.55$\AA$ with EP=3.89\,eV (right). Both lines have been computed at a fixed \feh$= -2.0$.}\label{fig2}
\end{figure*}
%\begin{figure*}
%\plotone{4070.77_3.24_FeI_GCOG.pdf}
%\caption{GCOG of EWs as a function of two-dimensional stellar parameters of the \fei\ line at 4070.77$\AA$, with EP=3.24\,eV. [Fe/H] is kept constant at $-2.00$ for demonstration purposes and to avoid cluttering the plot.\label{fig2}}
%\end{figure*}

%\begin{figure*}
%\plotone{6247.55_3.89_FeII_GCOG.pdf}
%\caption{Same as Figure\,\ref{fig2} but for the \feii\ line at %6247.55$\AA$ with EP=3.89\,eV. \label{fig3}}
%\end{figure*}

\subsection{Key Modules}\label{ssec:modules}
Below we describe in detail the key components and modules of the \texttt{LOTUS} functionalities. An illustrated flow diagram of the workings of the code is also shown in Figure\,\ref{fig:flowchart}, demonstrating the connection between the different modules.

\subsubsection{EW Interpolation  Module}\label{sect:ew_interp}
Traditional curve of growth (COG) methods to derive chemical abundances usually employ simplistic models, typically computed at   
%do not take into account the inter-dependencies of the EW of the lines on stellar parameters between 
fixed \teff, \logg, \feh\ and \vt\ for each line. For some lines, however, EW can have strong dependencies on multiple atmospheric stellar parameters in a given parameter space. Figure\,\ref{fig2} demonstrate these dependencies, where EW variation from our computed NLTE grid are shown as a function of \teff\ \logg\ and \vt\ for the Fe I line at 4070.768\,{\AA} (left panel), and the Fe II
lines at 6247.545\,{\AA} (right panel), respectively. The metallicity for both plots has been fixed at [Fe/H]\,$=-2.00$ for demonstration purposes. %\sout{It is clear that the EWs vary arbitrarily in the \teff-\logg\ space with resonance peaks showing throughout the grid at different parameters while demonstrating a smoother dependence in the \teff-\vt\ parameter space for the \fei\ line at 4070\,{\AA}. The \feii\ line at 6247\,{\AA}, however, shows smoother dependence on \teff\ and \logg\ for lower \teff\ and higher \logg\ values, with stronger resonance peaks showing at higher \teff\ and lower \logg.} 
We observe that at a fixed \feh, increasing \teff\ will decrease EW of the \fei\ line, while EW being mostly insensitive to \logg\ variation; For \feii\ lines, however, increasing \teff\ will increase EW while increasing \logg\ will decrease EW. For both lines, EW will increase as \vt\ increases, as \vt\ delays the saturation by spreading absorption into a wider wavelength band. These EW-stellar parameter dependencies have been qualitatively proven in Chapter 17.5 of \citet{Hubeny2014}. 
Thus, to take into account such important dependencies in our codes, which can affect our derived abundances and thus our stellar parameter derivation, we compute a Generalized Curve of Growth (GCOG) for each \fei\ and \feii\ line in our linelist. We define the GCOG for any given line $i$ as:
\begin{equation*}\label{eqn:1}
    GCOG_i\mathpunct{:}\left(T_{\mathrm{eff}}, \mathrm{log}\textit{g}, \nu_{mic}, \mathrm{EW}\right)\mapsto A(\mathrm{Fe})_i
\end{equation*}

A GCOG is thus a generalization of the well-known COG onto a 
higher dimensional mapping of EW $\mapsto$ A(Fe)
Alternatively, COG$_i$ can be obtained by fixing \teff\, \logg\, and \vt\ for a given GCOG. 
%\bf Figure\,\ref{fig2} demonstrates such dependencies.
 GCOGs have been implemented in previous studies such as \citet{Osorio2015} and \citet{Boeche2016}. We thus provide interpolated GCOGs which are stored as libraries in our code for every line in our linelist (see Section\,\ref{sec:linelist}).The GCOGs are then utilize in the optimization module afterward to derive the \fei\ and \feii\ abundances. 

\subsection{Abundance Determination module} \label{ssec:abund_module}
To derive the \fei\ and \feii\ abundances, we fit multivariate polynomials to the GCOG of each line, by deriving the relationship between the iron abundance of the line with \teff, \logg, \vt\ and EW, denoted as $A\left(\mathrm{Fe}\right) = P \left( T_{\mathrm{eff}}, \mathrm{log}\textit{g}, \nu_{mic}, \mathrm{EW} \right)$. The detailed form of $P$ can be written as:
\begin{equation}\label{eqn:2}
    \sum_{i_1+i_2+i_3+i_4\leq n} a_{i_1i_2i_3i_4}\left( T_{\mathrm{eff}} \right)^{i_1}\left(\mathrm{log}\textit{g}\right)^{i_2}\left(
    v_{mic} \right)^{i_3}\left(\mathrm{EW}\right)^{i_4}
\end{equation}
where $i_m$ ($m=1,2,3,4$) is the polynomial power of each of the 3 stellar parameters (\teff, \logg\ and \vt) as well as EW. %(In this case A(Fe) is equal to the input [Fe/H]). 
Possible values for $i_m$ are positive integers, where the sum of all the powers should less than or equal to the power degree of the polynomial $n$. $a_{i_1i_2i_3i_4}$ is the product of the polynomial coefficients of all the variables. The criteria for choosing $n$ depends on the behavior of theoretical GCOG models within a range of stellar parameters and is described in more details in Section\,\ref{sssec:select_interpolator} below. 

\begin{deluxetable}{ccc}
\tabletypesize{\footnotesize}
\tablecolumns{2} 
\tablecaption{\label{tab:interpolationvstype}Interpolation intervals chosen for different stellar types considered in \lotus.} 
\tablehead{\colhead{stellar parameter} & \colhead{types} & \colhead{intervals}}
\startdata
T$_{\textrm{eff}}$ (K) & K & [4000, 5200] \\
& G & [5200, 6000] \\
& F & [6000, 6850] \\
\hline
log$\textit{g}$ & supergiant & [0.0, 0.5] \\
& giant & [0.5, 3.0] \\
& subgiant & [3.0, 4.0] \\
& dwarf & [4.0, 5.0] \\
\hline
[Fe/H] & very metal poor & [$-3.5$, $-2.0$] \\
& metal poor & [$-2.0$, $-0.5$] \\
& metal rich & [$-0.5$, 0.5] \\
\enddata 
\end{deluxetable}

\subsubsection{Selection of Interpolator}\label{sssec:select_interpolator}
We choose the linelist utilized to derive stellar parameters to be dependent on the spectral type of the star, as some lines can be blended or strong in cool high metallicity stars, whereas these same lines could also be blend-free and weaker for hotter, metal-poor stars. Therefore, the multivariate polynomial functions defined in Section\,\ref{ssec:abund_module} used to derive the abundances determined from \fei\ and \feii\ lines depend on the choice of linelist for each spectral type. We thus pay special attention to the choice of interpolator by identifying the best selection of linelist per spectral type (or stellar parameters). Therefore, instead of interpolating in the whole parameter space for all the lines, we pre-define that the GCOGs fit only the abundances determined from the lines we pre-selected and chose to use within tested intervals of stellar parameters, according to an initial guess of spectral types that the users can insert as an input in the code. Our choice of the intervals of different stellar parameters for each spectral types are listed in Table \ref{tab:interpolationvstype}. For example, if an initial guess is chosen such that the target star is a metal-poor G giant, the multivariate polynomial will fit the abundance of each line versus other atmospheric parameters using the grid points falling into the range of \teff\ from 5200\,K to 6000\,K, \logg\ from 0.0 to 3.0 and [Fe/H] from $-2.0$ to $-0.5$. 

%The choice of lines for each stellar parameter interval of each spectral type was thoroughly tested.    
%Since the determination of iron abundance depends on the interpolation, we can't trust the accuracy of every line model. In order to select lines which have convincing interpolated result, we need to quantify the performance of interpolation line by line and star by star. In addition, we design a module for users to select convincing lines according to their spectroscopic analysis of EWs.
%First, we calculate the residuals of the theoretical EWs in each grid stellar atmosphere model and the interpolated EWs from multivariate polynomials with certain n, given the same stellar parameters as those in corresponding theoretical one. However we can't directly obtain EW by just plug stellar parameters into interpolated polynomial because we need solve one more non-linear equation as this:
%\begin{equation}
%    \sum_{i_1+i_2+i_3+i_4\leq n} c_{i_1i_2i_3i_4}\left(\textrm{EW}\right)^{i_4} = 0
%\end{equation}
%n is the degree of polynomial with same value mentioned in \ref{sssec:abund_module}. $c_{i_1i_2i_3i_4}$ is the coefficient obtained after plugging other stellar parameters into the interpolated model. We solve this equation via the algorithm depends on the eigenvalues of the companion matrix of this polynomial \citep{horn_johnson_2012}, which has been implemented in \texttt{NUMPY}.
In order to choose the best interpolator per spectral type (i.e., the optimal $n$) among several multivariate polynomials as defined in Equation\,\ref{eqn:2}, we use a Bayesian Information Criteria (BIC) to select the best polynomial function for each line based on the mean residual differences between the theoretical EWs at the nodes of our pre-computed NLTE grid (computed at a combination of \teff, \logg, \feh, and \vt), to the EWs interpolated at these parameters. 
We assume that the optimal model is among the polynomials with degree=2,3,4,5. The BIC was calculated such as:
\begin{equation}
    \mathrm{BIC} = n\, \mathrm{ln}\left(\frac{\mathrm{SSR}}{n}\right) + k\,\mathrm{ln}(n)
\end{equation}
where SSR is the sum of the squared residuals between the theoretical EWs and the interpolated EWs from models at specific nodes of the grid, and
%We collect the residuals of all nodes of the grid library. 
$n$ is sample size (here we choose $n=4000$ nodes in each spectral type interval). $k$ is a free parameter (including coefficients and slope intercept), which can be calculated as k=$C_{d+4}^{d}$, where $d$ is the degree of the polynomial, and 4 is chosen as the number of dependent variables for each interpolated EW (namely, \teff, \logg, [Fe/H] and \vt). 

\begin{figure*}[ht!]
\plotone{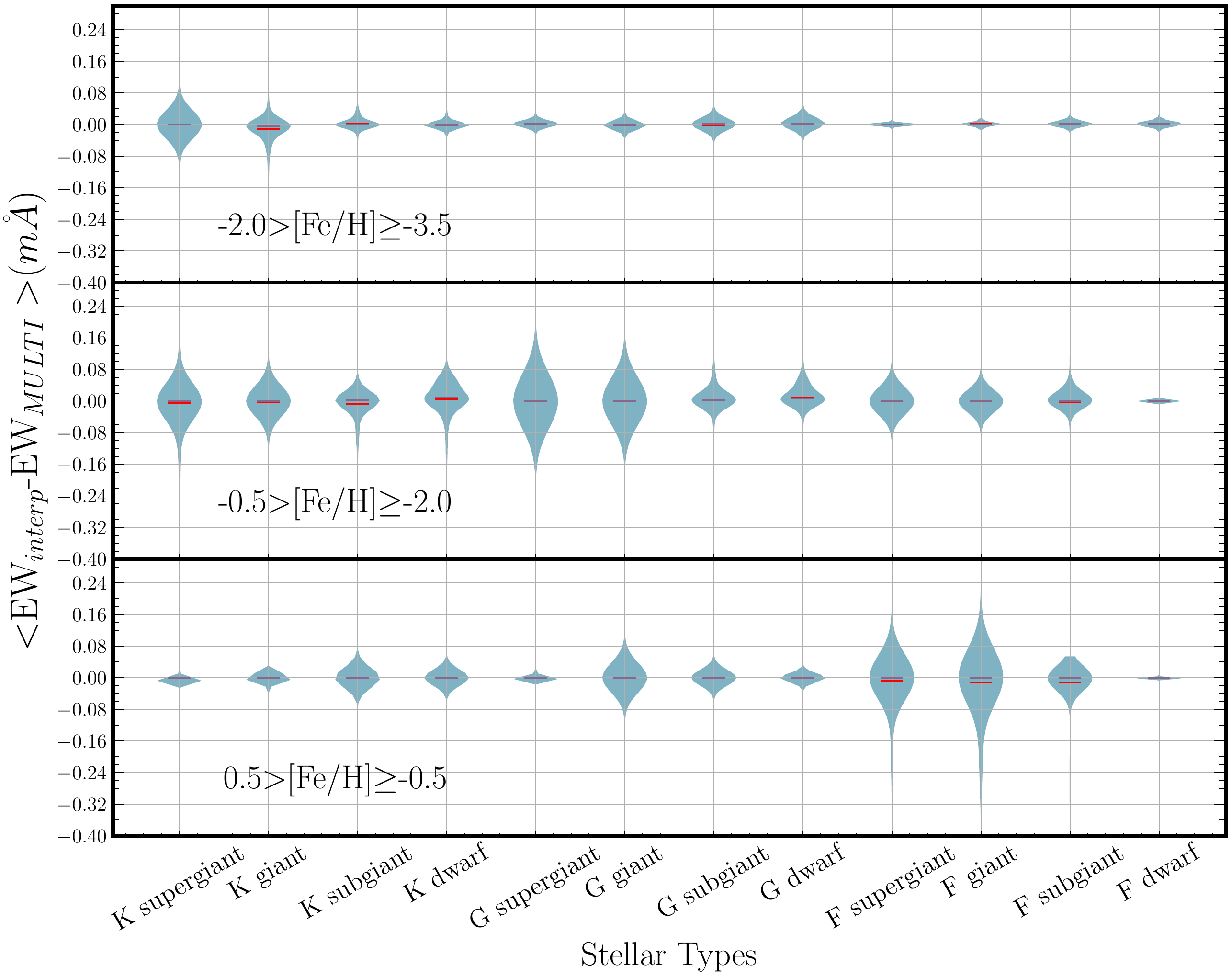}
\caption{Differences (in m{\AA}) between the interpolated EWs (using interpolators chosen at the lowest BIC values) and the EWs from NLTE theoretical calculations computed directly using \texttt{MULTI2.3}. Each violin plot shows the distribution of the means of such differences for a random 4000 nodes within the spectral type interval of the pre-computed EW library. Red lines are marked as the means of the distributions while purple lines are for the medians. 
%The outliers for each stellar spectral type are not shown in the plots, however  The numbers of outliers are less than 11. 
%The absolute value of difference for each stellar spectral type can be limited within 1 ignoring those outliers. 
The differences are shown for different metallicity ranges as indicated in each panel. \label{fig:4}}
\end{figure*}

We thus chose the best interpolator for each spectral type interval as that with the lowest BIC value. In Fig. \ref{fig:4}, we show the differences between the interpolated EW using the lowest BIC interpolator as compared to the (non-interpolated) EW directly from \texttt{MULTI2.3}, categorized as a function of stellar spectral types (x-axis) and metalicities (different panels).  We observe that there exist no clear dependencies between differences and spectral types and metalicities. The average of differences is within 0.01 for all stellar spectral types and metalicities. We pre-define the default interpolated EW uncertainty (also a threshold for the lowest acceptable BIC) in \lotus\ as 6\,m$\AA$ for each line. However, we also design the code to allow users to input their acceptable uncertainty limits, in which case 
\lotus\ will drop a given \fei\ or \feii\ from the linelist if the computed BIC is larger than this uncertainty, and will subsequently be excluded from the lines used to derive the stellar parameters in the optimization module.

\subsubsection{EP Cut-off}
The NLTE abundances derived from low excitation potential \fei\ can yield larger abundances as compared with those derived from high-excitation \fei\ and \feii\ lines, especially using 1D atmospheric models \citep{Amarsi2016}. Such differences can reach up to 0.45\,dex for some \fei\ lines, and can affect our derived stellar parameters as documented in several studies including \citet{Bergemann2012, Lind2012} as well as others. We thus follow previous literature studies by introducing low-excitation potential cut-offs for \fei\ lines with EP(\fei)\,$<$\,2.7ev/2.5ev/2.0\,eV for stars with [Fe/H]$<-0.5$, depending on the number of \fei\ and \feii\ lines measured in the stars and the optimization convergence criteria (see Section\,\ref{sect:optimization}); We conduct no cut-offs for stars with [Fe/H]$>-0.5$.  

\subsection{Optimization Module}\label{sect:optimization}
Once reliable linelist and interpolators per spectral types of input target stars have been chosen, as explained in Section\,\ref{ssec:abund_module} above, the derived iron abundances from \fei\ and \feii\ lines are fed into the optimization module with an initial guess of the general type of the target star. We note that users don't need to specify an initial guess of each stellar parameter, as it suffices to only chose an initial guess of the spectral type as an input to \lotus, which is then assigned the corresponding interpolation parameters interval as indicated in Table\,\ref{tab:interpolationvstype}.

\subsubsection{Optimization Conditions} \label{sssec:2.6.1}
The general principle of optimization is to adjust stellar parameters iteratively to derive optimal combinations of stellar parameters that can satisfy the following three conditions: (i) excitation equilibrium, or minimizing the trend (i.e., slope) of the \fei\ abundances as a function of excitation potential EP, (ii) ensuring ionization equilibrium or minimizing the differences between the abundances derived from the \fei\ and \feii\ lines, and (iii) minimizing the trend (i.e., slope) between the abundances derived from the \fei\ lines versus the reduced equivalent widths, REW\,=\,$\log(\mathrm{EW}/\lambda)$.

Therefore, we derive \teff\ by ensuring excitation equilibrium, \logg\ by ensuring ionization equilibrium, and \vt\ by minimizing the trend between \fei\ abundances versus line strength (REW). \feh\ was then determined by averaging the abundances derived from the \fei\ and \feii\ abundances.

\begin{figure*}
\epsscale{1.0}
\plotone{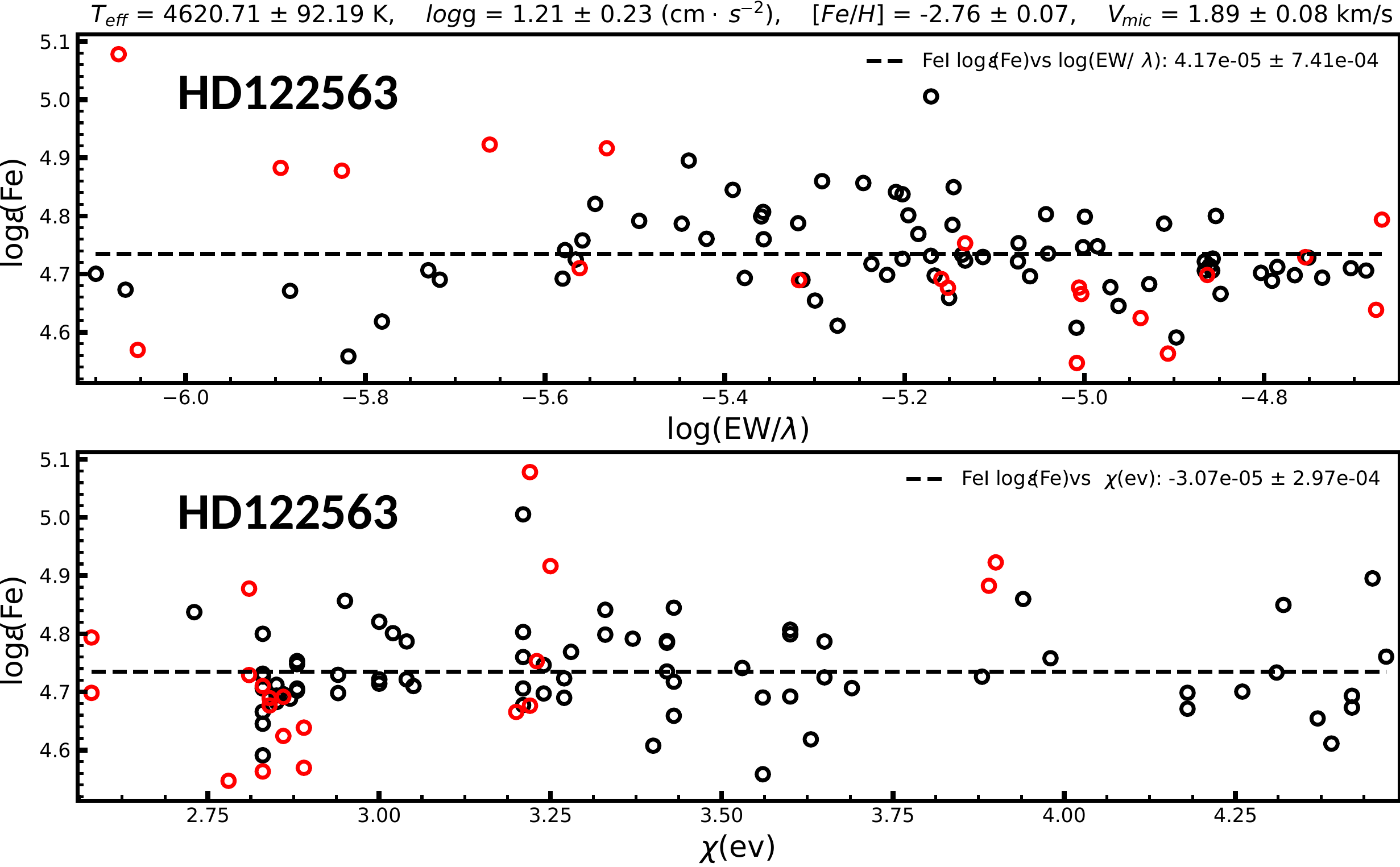}
\caption{Iron abundances determined from \fei\ (black circles) and \feii\ (red circles) line versus reduced equivalent widths (upper panel) and excitation potential energies $\chi$ (lower panel) for selected lines in the metal-poor benchmark star HD122563. The parameters indicated on the top are the optimal values derived using the global minimization (differential evolution) algorithm in \lotus. Dotted lines are the best linear fits to the \fei\ lines in each panel. The labels at the upper left corner of each panel are the slopes of the fits and their corresponding standard deviations.\label{fig:HD122563_equilibria}}
\end{figure*}

\subsubsection{Targeted Optimized Object Function}
In order to find the best combination of stellar parameters that satisfy the optimization conditions defined in Section\,\ref{sssec:2.6.1} within our parameter grid, we first combine these conditions into an object function $\mathcal{F}$, such as:
\begin{equation}\label{eq:6}
    \mathcal{F} = \left(\frac{s_{\chi,1}}{\sigma_{\chi,1}}\right)^2 + \left(\frac{s_{REW,1}}{\sigma_{REW,1}}\right)^2 + \left(\frac{\bar{A_1} - \bar{A_2}}{\sigma_{1-2}}\right)^2
\end{equation}
where the subscripts "1" and "2" correspond to \fei\ and \feii, respectively. $s_{\chi,1}$ is the slope between \fei\ abundances and EP, and $\sigma_{\chi,1}$ is its uncertainty, $s_{REW,1}$ is the slope between the \fei\ abundances and REW, and $\sigma_{REW,1}$ is its uncertainty. $\bar{A_1}$ and $\bar{A_2}$  are the mean abundances for \fei\ and \feii, respectively, while $\sigma_{1-2}$ is the standard deviation of the differences between $\bar{A_1}$ and $\bar{A_2}$. A converged solution is obtained at the combination of \teff, \logg, \feh\ and \vt, which minimizes $\mathcal{F}$ globally.

\subsubsection{Global minimization}\label{sec:glob_optim}
To find the global minimization fit parameters within our grid, we adopt a differential evolution algorithm implementing a global minimization search. The goal is to minimize $\mathcal{F}$ by starting with an initial population of candidate solutions, which are iteratively improved by retaining the fittest solutions that yield a lower $\mathcal{F}$ values, until convergence for the best-fit parameters is met \citep{Storn1997}. The advantage of a differential evolution algorithm is that it has the benefit of handling nonlinear and non-differentiable multi-dimensional objective functions while requiring few control parameters to steer the minimization.
%propose iteratively several candidates which are randomly distributed within the desired bound. These candidates are moved around the parameter spaces according to simple formulae which combine the relative positions of these candidates. If the new position provide minimum value of the objective function among current population of candidates, it is accepted and forms part of the population. Otherwise, this new position discarded. This process continues iteratively until one or more termination criteria is/are met. It can extend a large searching areas and do not require the gradients of the objective function towards the parameters. . 
For our code, we select 100 initial populations (sets of solutions) with a combined rate of 0.3, which is a crossover probability that depends on how fast the algorithm moves to the next generation of populations. We try different weights between 0.8 and 1.2, to maintain a proper searching radius, at the same time making sure it does not slow down the convergence speed. We adopt $\mathcal{F} \sim 10^{-5}$ as the absolute tolerance for each iteration. After each iteration, we perform a 3$\sigma$-clipping on the abundances determined from the \fei\ and \feii\ lines to remove outliers. We adopt 3 iterations of outlier removing in total as a default, however, this number can be defined by the users depending on the quality of their EW measurements. An example of HD122563 with derived abundances of \fei\ and \feii\ versus REW and EP at optimal atmospheric parameters is shown in Figure \ref{fig:HD122563_equilibria}.

\subsection{Uncertainty Estimation Module}\label{sssec:uncert_estimation}
\begin{figure*}[ht!]
\plotone{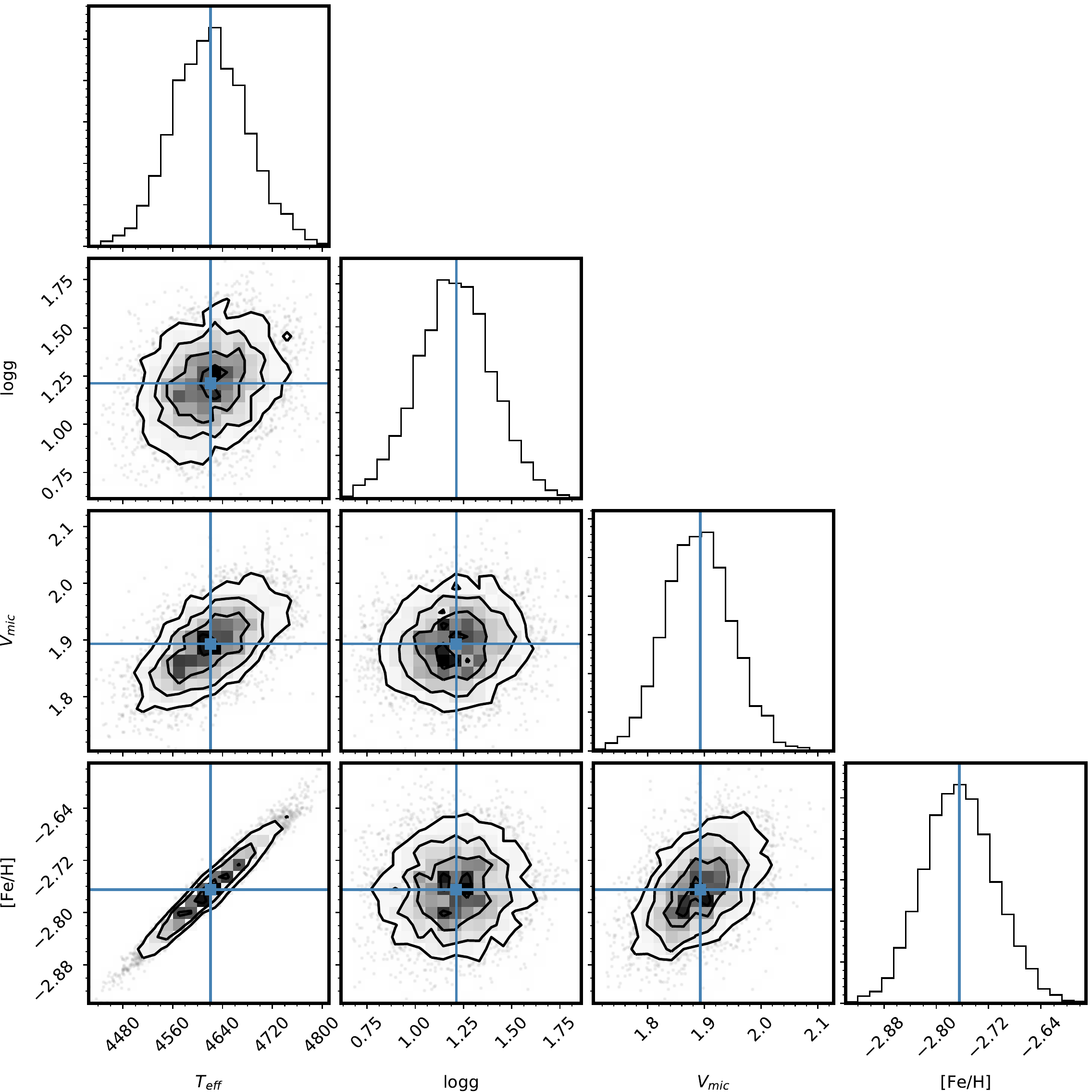}
\caption{Marginalized posterior distributions of the stellar parameters of HD122563 based on the log-likelihood function in Equation\,\ref{eq:9}. 1$\sigma$, 2$\sigma$ and 3$\sigma$ uncertainty region are denoted as solid contours in the two dimensional plot. The histograms show the marginalized posterior distributions for each parameter, respectively. Blue intersecting lines show the values obtained from our global optimization module. \label{fig5}}
\end{figure*}

Since $\mathcal{F}$ is a function of the stellar parameters, it can be described as $\mathcal{F}(\theta) = \mathcal{F}(T_{\mathrm{eff}}, log\mathrm{g}, $\vt$)$. We can keep track of how $\mathcal{F}$ changes along the perturbations around the optimal solution $\theta^*$=($T_{\mathrm{eff}}^*$, log$\textit{g}^*$, \vt$^{*}$), $T_{\mathrm{eff}}^*$, log$\textit{g}^*$ and \vt$^{*}$ are the parameters at the optimal solution. Indeed, $\mathcal{F}$ resembles a likelihood function, which can be used to estimate the uncertainties on each of our derived stellar parameters from the Hessian Matrix such as:
\begin{equation}
    SE(\theta^*) =  diag(\sqrt{\mathcal{H}^{-1}(\theta^*)})
\end{equation}
where the Hessian Matrix for our objective function can be written as:
\begin{equation}
    \mathcal{H}(\theta) = \frac{\partial}{\partial \theta_i \partial \theta_j}\mathcal{F}(\theta), 1\leq i,j \leq 3
\end{equation}
For [Fe/H], we adopt the standard deviation of the \fei\ lines as the uncertainty.

However, in the above uncertainty framework, one strong assumption is that the standard errors obtained are symmetric around the mean values. This is only true if the probability distribution function of the derived parameters is a normal distribution. Normally, this is not the case. We, therefore, use a Bayesian framework to robustly determine the uncertainties on our derived stellar parameters. First, we construct a  log-likelihood function using the same terms as in Equation\,\ref{eq:6} such as:
\begin{equation}\label{eq:9}
    log(\mathcal{L}) = -\frac{1}{2}\sum_{i\leq3}\left[\left(\frac{s_i}{n_i}\right)^2+\log(2\pi n_i)\right]
\end{equation}
where $n_i$ can be written as:
\begin{equation}
    n_i = \sigma_i^2 + f^2s_i^2
\end{equation}
where $s_i$ represent the slopes $s_{\chi,1}$, $s_{REW,1}$ and $\bar{A_1} - \bar{A_2}$ individually, while $\sigma_i$ are their corresponding standard deviations. Introducing $f$ will compensate the underestimation of the variance of each parameter assuming such an additional term is proportional to the model itself. 

We then perform the Slice Sampling algorithm \citep{Neal2003}, which is a type of Markov Chain Monte Carlo (MCMC) implemented in \texttt{PyMC3}, to complete the estimation of the posterior probability of each parameter. This method adjusts the step size automatically on every proposed candidate to match the profile of the posterior distribution without the need to choose a transition function. Such a framework ensures higher efficiency as compared to classical MCMC algorithms, such as Metropolis and Gibbs \citep{Metro1949, Geman1953}. For all stellar parameters, we derive a normal distribution around the mean optimized value obtained from the global minimization module described in Section\,\ref{sec:glob_optim}, and a standard deviation determined from the standard errors of the Hessian Matrix.

We use the logarithmic of $f$ as an input variable to the MCMC sampler with a uniform distribution ranging from $-$10 to 1. In the sampling process, we consider 4 running chains with each containing 1000 draw steps and 200 tunes steps. Figure \ref{fig5} shows the output from our posterior distribution of the stellar parameters and their uncertainty regions for a benchmark metal-poor star, HD122563. On the same plot, we also show the optimized parameters derived using our global optimization module (blue lines intersection), which agree very well.

\section{Testing \texttt{LOTUS}\label{sec:test}}
We test \texttt{LOTUS} by deriving the atmospheric stellar parameters of benchmark stars with well constrained non-spectroscopic and fundamental stellar parameters, to quantify the precision of our derived parameters and their uncertainties. Below we present the results of these tests. We list the derived stellar parameters (\teff, \logg, \feh\ and \vt) using \lotus\ in LTE and NLTE of all the stars considered from Section \ref{sssec:3.2.1} to \ref{sssec:3.2.4} below, the sources of the EW, the number of \fei\ and \feii\ lines, and the excitation potential cut-offs for each star in Table \ref{tab:derived_parameters}. 

\begin{figure*}[ht!]
\plotone{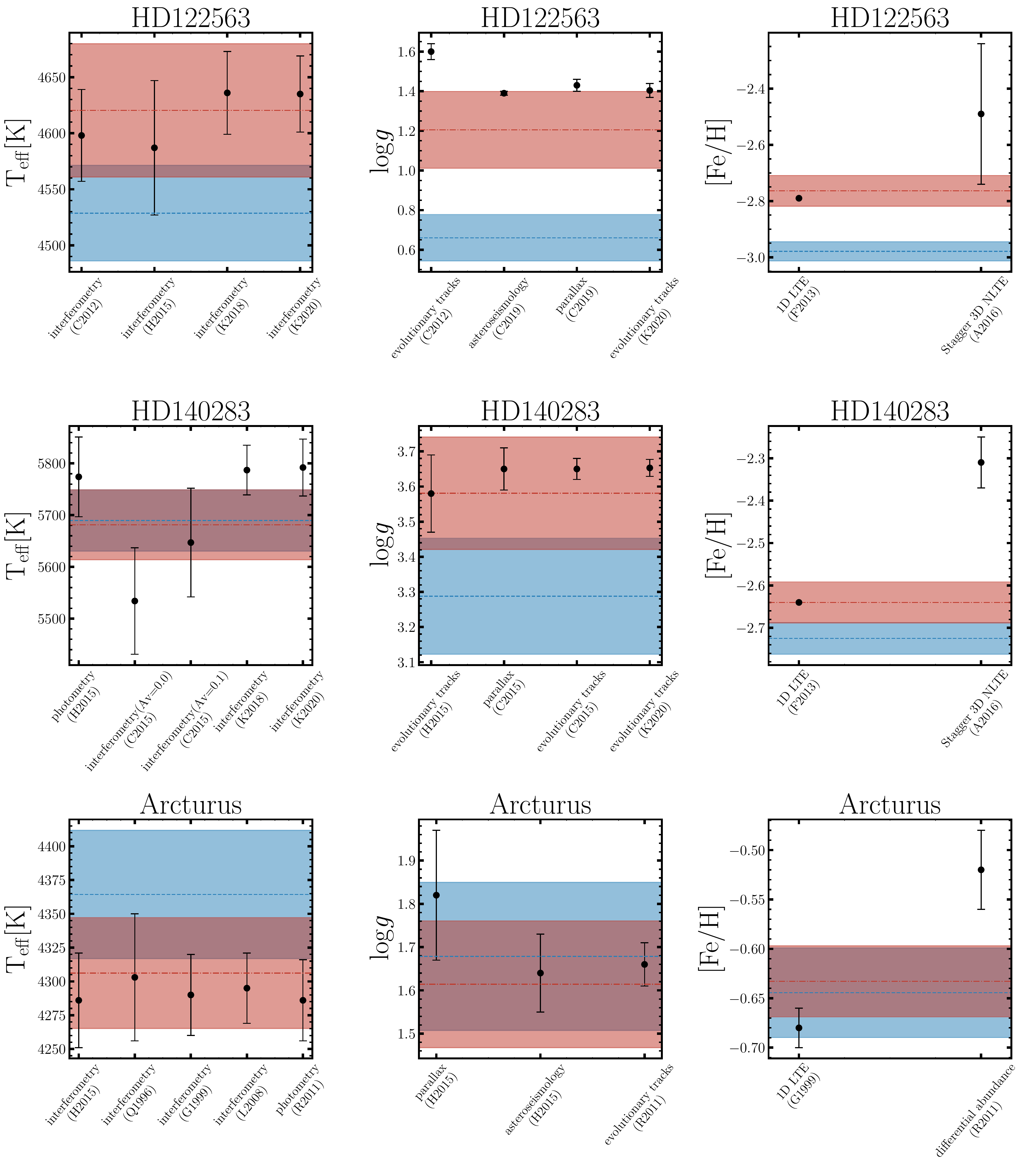}
\caption{Comparison of our derived stellar parameters, \teff, \logg\ and [Fe/H] using \lotus\ to reference literature values derived non-spectroscopically (for \teff\ and \logg) for HD122564, HD140283 and Arcturus.  The mean NLTE values (red dashed lines) with their 1$\sigma$ confidence intervals (red shadow areas) and our mean LTE values (blue dashed lines) with their 1$\sigma$ confidence intervals (blue shadow areas), compared to to reference stellar parameters (black dots with error bars) for references on the x-axis. References are as follows: Q1996...\citet{Quirrenbach1996}; G1999...\citet{Griffin1999}; L2008...\citet{Lacour2008};R2011...\citet{Ramirez2011};C2012...\citet{Creevey2012};F2013...\citet{Frebel2013};H2015...\citet{Heiter2015};C2015...\citet{Creevey2015};A2016...\citet{Amarsi2016};K2018...\citet{Karo2018};C2019...\citet{Creevey2019};K2020...\citet{Karo2020}. \label{HD12&HD14&Arc}}
\end{figure*}

\begin{figure*}[ht!]
\plotone{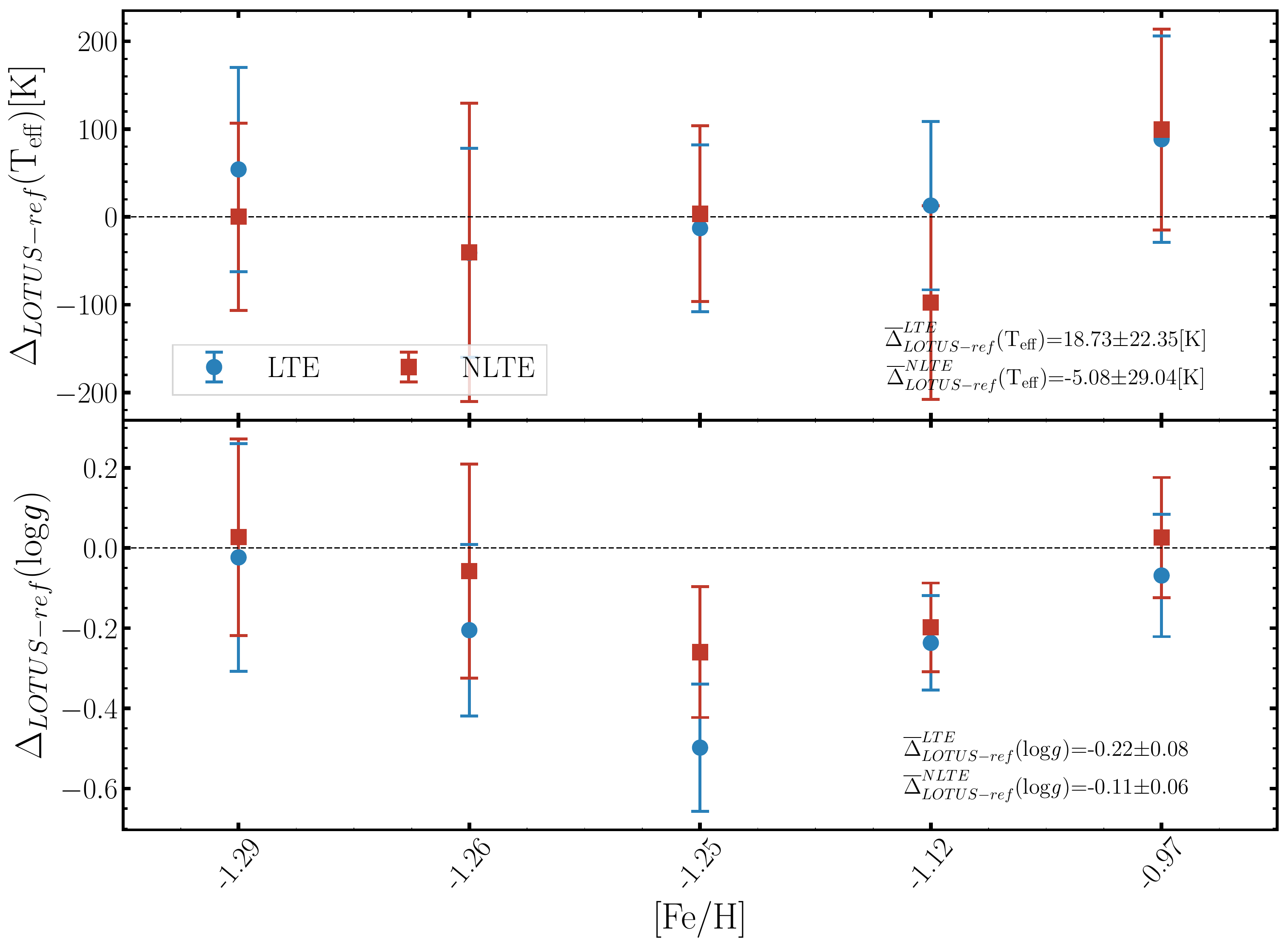}
\caption{Differences in \logg\ and \teff\ between values derived using \lotus\ versus reference values from 
\citet{Hawkins2016} for five GES metal-poor stars in their sample. Blue circles are LTE differences, while red squares are NLTE. Dashed lines in both panels are the zero baselines. Average systematic offsets are included in the text of each panel.\label{GES-MP_v0}}.
\end{figure*}

%As a pioneer of establishing set of calibrators that can test different spectroscopic analysis tools with non-spectroscopic analysis, Gaia FGK benchmark stars (GBS) sample provided independent measurement of T$_{\mathrm{eff}}$ and \logg, via the star’s angular diameter ($\theta_{LD}$) and bolometric flux ($F_{bol}$) \citep{Heiter2015}, following with a plenty of candidates inside the region of low metalicity ($-1.3<$[Fe/H]$<-1.0$) \citep{Hawkins2016}(H2016). Furthermore, \citet{Worley2020}(W2020) extended stellar spectroscopic datasets combined with precise asterseismological measurements from K2. Due to the high precision of surface gravity determined from asterseismology (around $0.01$ dex),  we can decrease the degeneracy of stellar parameters to equivalent width and therefore constrain T$_{\mathrm{eff}}$, $\xi_t$ and [Fe/H] better. Recently, \citet{Karo2020, Karo2021a, Karo2021b} published a series of paper to determine fundamental stellar parameters from CHARA interferometry for several metal stars, metal rich subgiants and dwarfs, which can be a homogeneous benchmark set. 

\subsection{HD122563, HD140283, and Arcturus\label{sssec:3.2.1}}
The two metal-poor standard stars HD122563, HD140283 and the benchmark giant Arcturus have been widely analyzed independently in the literature. Their stellar parameters have therefore been derived spectroscopically (using 1D, 3D, LTE and NLTE assumptions, or a combination of each), as well as using non-spectroscopic methods and fundamental equations (e.g., using photometric calibrations, asteroseismology, or interferometric angular diameters). All three stars are dubbed benchmark stars which have been selected as comparison standards for the largest stellar surveys, especially in the $Gaia$-ESO spectroscopic survey \citep{Jofre2014,Heiter2015}. We adopt EW measurements of each benchmark star from literature publications as follows: HD122563 \citep{Frebel2013}, HD140283 \citep{Frebel2013}, Arcturus \citep{Jofre2014}. \citet{Frebel2013} measured their EW from high-resolution spectroscopic observations ($R\sim$35,000 in the blue and $\sim$28,000 in the red) of HD122563 and HD140283, obtained with the MIKE spectrograph \citep{Bernstein2003} on the Magellan-Clay telescope at Las Campanas. A $R\geq$70,000 spectrum was used to derive EW for Arcturus from the VLT spectrum \citep{Jofre2014}. We derive the stellar parameters of each star and their uncertainties using \texttt{LOTUS}. We then compare our NLTE and LTE parameters with several independent determinations from the literature, as shown in Figure \ref{HD12&HD14&Arc}.

\paragraph{HD122563} We derive \teff\,=\,4620$\pm$59\,K, \logg\,=\,1.21$\pm$0.19, \vt\,=\,1.89$\pm0.06$\,\kms\ and [Fe/H]\,=\,$-$2.76$\pm$0.05 in NLTE, and \teff\,=\,4528$\pm$42\,K , \logg\,=\,0.66$\pm$0.12, \vt\,=\,1.84$\pm0.03$\,\kms\ and [Fe/H]\,=\,$-$2.98$\pm$0.03. \citet{Creevey2012} derived \teff\,=\,4598$\pm$41\,K using an angular diameter observed with CHARA interferometry and Palomar interferometeric observations. \citet{Heiter2015} derived \teff\,=\,4587$\pm$60\,K based on the angular diameters and bolometric fluxe calibrations, and \logg\,=\,1.44$\pm$0.24 from fitting evolutionary tracks.
\citet{Karo2018} updated CHARA interferometry with more data and derived an effective temperature of \teff\,=\,4636$\pm$37\,K. Later in \citet{Karo2020}, they updated their data reduction pipeline and derived a new \teff\,=\,4635$\pm$34\,K. Our NLTE \teff\ derived for HD122563 agrees very well with \teff\ determined in these studies using interferometric angular diameters, within $\sim$30\,K, while the LTE \teff\ deviated by $\sim 100$\,K from the reference values.  

And for \logg\, results from the recent updated asteroseismic analysis are in agreement with the NLTE values in our work, which is close to the upper limit of 1$\sigma$ confidence interval.

For \logg, \citet{Creevey2012} used evolutionary track models to derive \logg\,=\,1.60$\pm$0.04 for HD122563, while
\citet{Creevey2019} utilized the Hertzsprung telescope (SONG network node) to accurately measure the surface gravity of HD122563 using asteroseismology for the first time, and derived \logg\,=\,1.39$\pm$0.01. In their paper they also compare to $Gaia$ DR2 parallax-based \logg\,=\,1.43$\pm$0.03, which shows high consistency between these methods. \citet{Karo2020} also used evolutionary track models to derive \logg\,=\,1.404$\pm$0.03 for the same star. Our \lotus\ \logg\,=\,1.22 in NLTE matches within $\sim 0.1$\,dex the asteroseismic and parallax \logg\ in \citet{Creevey2019}, whereas the LTE value is 0.6\,dex lower. This gives us strong confidence in \lotus' ability to derive NLTE surface gravities from spectroscopic observations. 

We then compare our NLTE and LTE \feh\  with those derived by \citet{Frebel2013} in 1D, LTE and \citet{Amarsi2016} in 3D,NLTE. Our NLTE \feh\ agrees with that determined in both studies within error bars, although \citet{Amarsi2016} derived a higher \feh\,=\,$-2.5$, which is likely due to considering 3D effects in their study. 
%Noticeably, \citet{Amarsi2016} predicted that their input parallax-based surface gravity \logg=1.1 is overestimated, which is within 1$\sigma$ of our NLTE result.

\paragraph{HD140283} 
We derive \teff\,=\,5681$\pm$67\,K , \logg\,=\,3.58$\pm$0.16, \vt\,=\,2.17$\pm$0.16 \kms\ and [Fe/H]\,=\,$-2.64\pm$0.05 in NLTE, and \teff\,=\,5689$\pm$59\,K , \logg\,=\,3.29$\pm$0.17, \vt\,=\,2.09$\pm$0.14 \kms\ and [Fe/H]\,=\,$-2.72\pm$0.04 in LTE.
\citet{Heiter2015} derived a mean \teff\,=\,5774$\pm$77\,K for the sub-giant in their $Gaia$-ESO benchmark sample from photometric calibrations. Additionally, \citet{Creevey2015} used the VEGA interferometer on CHARA to determine \teff\,=\,5534$\pm$103\,K (using $A_V$\,=\,0.0 mag) and \teff\,=5\,647$\pm$105\,K (using $A_V$\,=\,0.1 mag). \citet{Karo2018} similarly derived an updated \teff\,=\,5787$\pm$48\,K from additional interferometric data points, and afterwards updating their values to 5792$\pm$55\,K in \citet{Karo2020}. Our LTE and NLTE \teff\ for HD140283 agree with the interferometric values derived in \citet{Creevey2015} within 120\,K and \citet{Karo2018,Karo2020} within 80\,K. 

%\citet{Heiter2015},  reported it's \teff and \logg, they do not recommend use HD140283 as a reference star for \teff validation since there exists large discrepancy between photometric one and their value.  
\citet{Heiter2015} derived \logg\,=\,3.58$\pm$0.11 using fundamental relations and adopting independently derived parallax for the star. \citet{Creevey2015} also derived a mean \logg\,=\,3.69$\pm$0.03, similarly from a combination of parallax and evolutionary track models.  Both \logg\ determined by parallax methods in \citet{Heiter2015} and \citet{Creevey2015} agree well with our NLTE value to within 0.05\,dex. \citet{Karo2020} derived \logg\,=\,3.65$\pm$0.0 from evolutionary models. Our LTE values is, unexpectedly, 0.3\,dex lower. 
%They argued that their observations at visible wavelengths can avoid systematic errors arising by the unresolved size of stars at the near-infrared band in \citet{Creevey2015}. And both our \teff\ in NLTE and LTE agree well with this recent value.

Similar to HD122563, we compare our \feh\ with 1D,LTE and 3D,NLTE abundances in \citet{Frebel2013} and \citet{Amarsi2016}, respectively. Given 
that both studies derived their abundances by fixing \teff\ and \logg, while ours were derived simultaneously using the global optimization method adopted in \lotus, we warrant serious direct comparison with their values.
Nevertheless, our NLTE values agree very well with that derived in \citet{Frebel2013} within 1$\sigma$, whereas is 0.3\,dex lower as compared to \citet{Amarsi2016}.

\paragraph{Arcturus} 
We derive \teff\,=\,4306$\pm$41\,K, \logg\,=\,1.61$\pm$0.15, \vt\,=\,1.79$\pm$0.02 \kms\ and [Fe/H]\,=\,$-0.63\pm$0.04 in NLTE, and \teff\, =\,4364$\pm$47\,K , \logg\,=\,1.68$\pm$0.17 , \vt\,=\,1.76$\pm$0.02 \kms\ and [Fe/H]\,=\,$-0.64\pm$0.05 in LTE.
For \teff\, \citet{Quirrenbach1996} used the MkIII Optical Interferometer on Mt. Wilson to determine Arcturus' angular diameter and derived \teff\,=\,4303$\pm$47\,K. \citet{Griffin1999} collected all literature results of interferometry observation up to their study and derived \teff\,=\,4290$\pm$30\,K. \citet{Lacour2008} used the IOTA 3 telescope interferometer in the H band to derive \teff\,=\,4295$\pm$26 K. Additionally, \citet{Ramirez2011} fit theoretical spectral energy distributions (SED) from visible blue bands to mid-infrared to derive \teff\,=\,4286$\pm$30\,K. Finally, \citet{Heiter2015} derived a mean of 4274$\pm$83\,K from different interferometric measurements, even though they warranted against using Arcturus as a benchmark given the large dispersion they obtained. All interferometric \teff\ agree very well with our NLTE \teff\ to within $\sim 20$\,K. Our LTE value is, however, $\sim 80$\,K higher.

\citet{Ramirez2011} derived \logg\,=\,1.66$\pm$0.05 from HIPPARCOS parallax and isocrhone fitting, while 
\citet{Heiter2015} derived a fundamental \logg\,=\,1.64$\pm$0.09 based on seismic mass and \logg\,=\,1.82$\pm$0.15 from a compilation of parallax-based measurements. Our NLTE \logg\ agrees very well with both results, whereas our LTE values is $0.1$\,dex higher.

\citet{Griffin1999} derived \feh\,=\,$-0.68\pm$0.02 by comparing theoretical 1D, LTE SEDs with observed flux.  \citet{Ramirez2011} used a differential abundance analysis relative to the solar spectrum to derive [Fe/H]\,=\,$-0.52\pm$0.04. Our NLTE and LTE [Fe/H] are within 0.05\,dex from each other, and agree well with the values derived in \citet{Griffin1999} while being $\sim 0.1$\,dex as compared to differential abundance analysis in \citet{Ramirez2011}.

\begin{figure*}[ht!]
\plotone{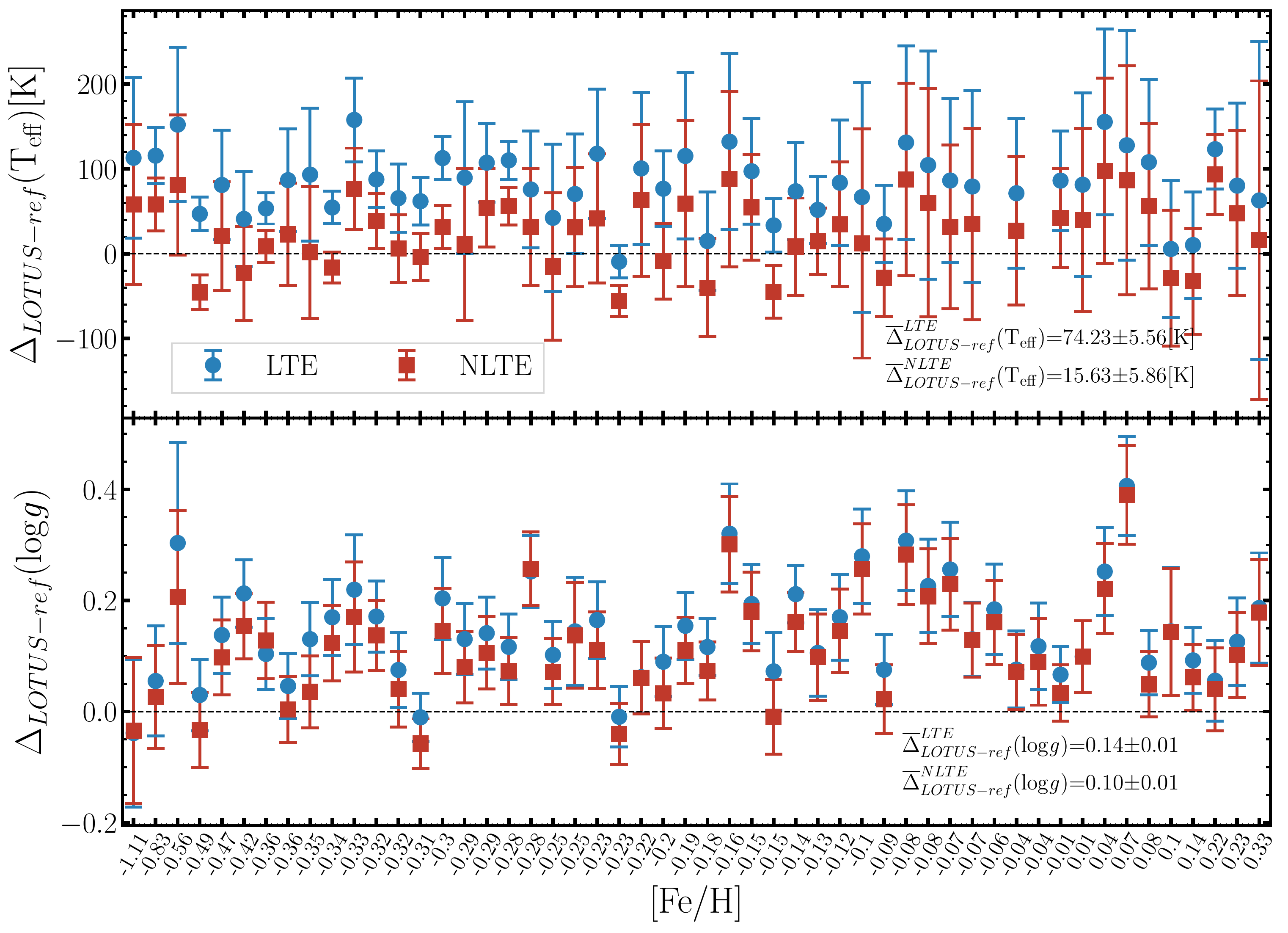}
\caption{Same as in Figure\,\ref{GES-MP_v0} for the GES-K2 sample stars from \citet{Worley2020}. \label{K2_v0}}
\end{figure*}

\begin{figure*}[ht!]
\plotone{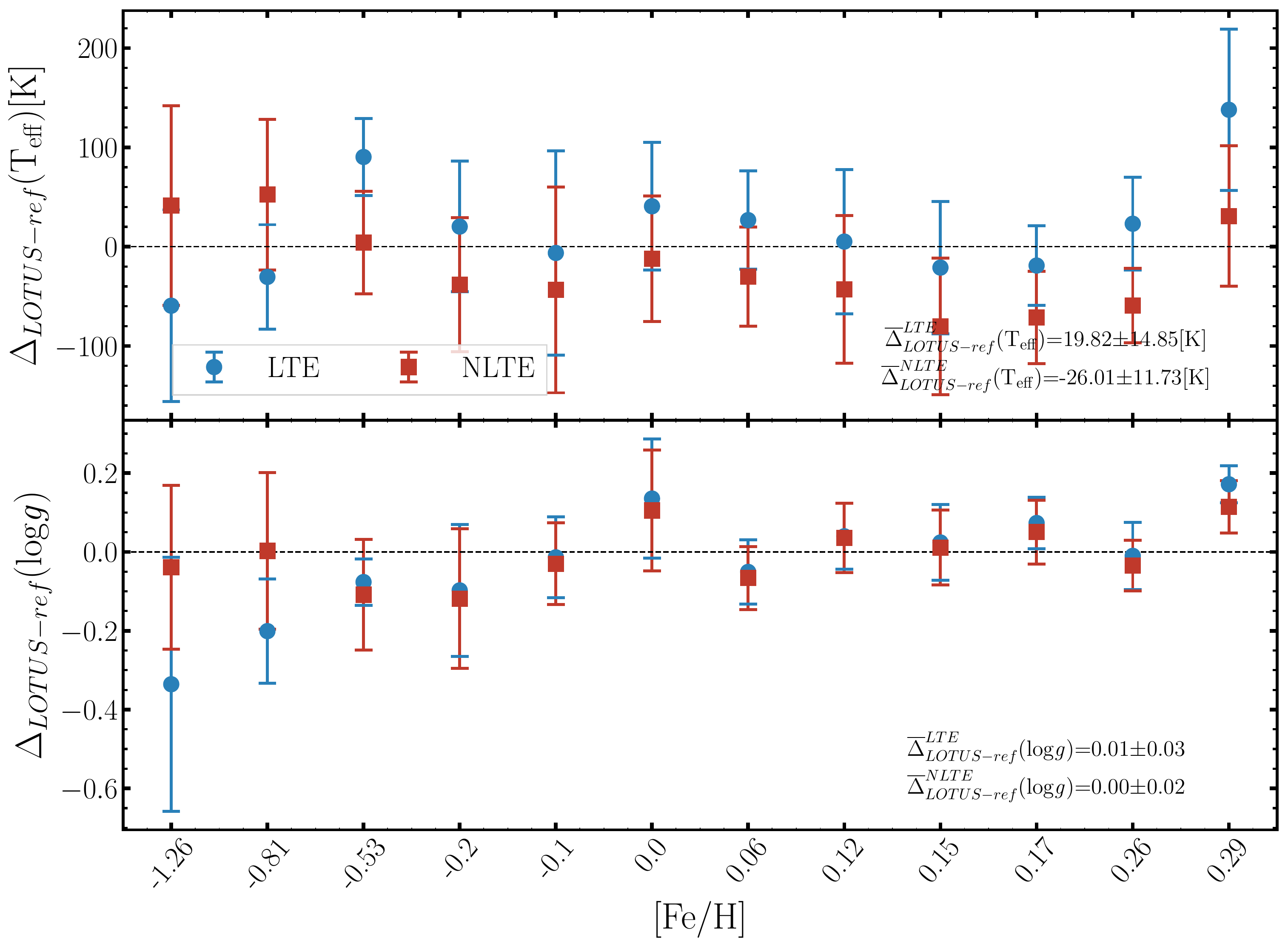}
\caption{Same as in Figure\,\ref{GES-MP_v0} for the CHARA sample stars from \citet{Karo2020, Karo2021a, Karo2021b}. \label{fig:CHARA_v0}}
\end{figure*}

\subsection{GES Metal-Poor Stars}\label{sssec:3.2.2}
We also test \texttt{LOTUS} on five metal-poor benchmark stars proposed by \citet{Hawkins2016} from the $Gaia$-ESO survey (GES). 
They determined \teff\ via the Infrared Flux Method (IRFM) \citep{Casagrande2011} using multi-band photometry and \logg\ by fitting to evolutionary stellar models. 
The \fei\ and \feii\ EWs for the sample stars were adopted from \citet{Hawkins2016}, who used different EW measurements from different pipelines from the GES survey to derive their parameters. We use their EPINARBO (EPI) EWs derived with the FAMA pipeline following \citet{Magrini2013}, as those were derived consistently. 
We compare our derived \teff\ and \logg\ using \texttt{LOTUS} to those in \citet{Hawkins2016} using the same \fei\ and \feii\ lines. We find that our NLTE parameters are on average within $5\pm 29$\,K from their \teff\ and $0.11\pm0.06$ for \logg\ on average, whereas our LTE parameters are within $18\pm22$\,K and $0.22\pm0.08$\,dex for \teff\ and \logg\ respectively. Star by star comparisons are shown in Figure\,\ref{GES-MP_v0}. 
Our NLTE parameters agree better with the non-spectroscopic parameters from \citet{Hawkins2016}, as compared to LTE.

A larger \logg\ dispersion is obtained for HD106038 though with [Fe/H]=$-1.25$ of $-0.25$ and $-0.45$, for both our NLTE and LTE results, respectively. With a 3D, Non-LTE analysis, and $Gaia$ EDR3 parallax, \citet{Giribaldi2021} derived \logg=4.29$\pm0.04$, which is in excellent with our NLTE  \logg=4.29$\pm$0.08. Additionally, only 35 \fei\ lines used by \citet{Hawkins2016} were used to derive the stellar parameters in \lotus\,  which decreases the accuracy of optimal values (see Section\,\ref{sec:caveats} for more details).

%We fine that on average our parameters derived with \texttt{LOTUS} are withindown the dispersion as compared to H2016 from 1.5-2.0$\sigma$ in LTE, to For HD106038, we find a 1.7$\sigma$ dispersion in \teff\ as compared to \teff\ a 3.4$\sigma$ for \logg\ as compared to H2016, while a 1.5$\sigma$ of the difference differs with \teff adopted in H2016 and a 2.0$\sigma$ disagreement of \logg with that adopted in H2016 for HD102200; a 1.8 $\sigma$ disagreement of \logg with that adopted in H2016 for HD201891. For NLTE, disagreement for all comparisons are close or smaller than 1$\sigma$. 

\subsection{GES-K2 Stars}\label{sssec:3.2.3}
We also test \lotus\ by deriving atmospheric stellar parameters of a sample of Kepler-2 (K2) star sample, which was also observed using a high-resolution UVES spectrograph on the VLT as part of the GES survey. \citet{Worley2020} combined the high-resolution spectroscopic observations from UVES, photometry, and precise asteroseismic data from K2 to derive self-consistent stellar parameters for these stars, which represent a good non-spectroscopic sample to compare our stellar parameters derived with \texttt{LOTUS} to.

We adopt the EW measurements from \citet{Worley2020} derived from high-resolution UVES spectroscopic observations. We chose a total of 52 stars from their sample. 
%However, in their paper, their result of two stars, CNAME 22000793-1203412 and CNAME 22032202-0829154, do not agree with other spectroscopic values and their \teff have large offset with IRFM ones. Therefore, we do not include this 2 stars in the final comparison. 
The majority of the stars in the sample are metal-rich with \feh$>-0.5$. Only six stars have $\feh<-0.5$. \citet{Worley2020} used the IRFM to determine photometrically calibrated \teff\ for all their stars following \citet{Casagrande2010,Casagrande2011}.

%In order to compare our fitting with non-spectroscopic values as much as possible, we take them as benchmarked \teff. For \logg, we take their final seismic determined ones.
The average uncertainties reported in \citet{Worley2020} for \teff\ and \logg\ of the GES-K2 stars are $\pm65$\,K and $\pm0.02$, respectively. We note that these values are close to our parameter grid resolution.
%These uncertainties are close to our parameter grid resolution while the uncertainty of reference \teff is far smaller than our grid resolution. 
We derive the stellar parameters for the 52 GES-K2 stars using \texttt{LOTUS}. Our results for \logg\ and \teff\ as compared to \citet{Worley2020} are shown in Figure\,\ref{K2_v0} as a function of \feh.
We find that the average differences between the \lotus\ parameters are $\bar{\Delta}$\teff\,=\,74$\pm$6\,K and $\bar{\Delta}$\logg\,=\,0.14$\pm$0.01 in LTE, and $\bar{\Delta}$\teff\,=\,16$\pm$6\,K, $\bar{\Delta}$\logg\,=\,0.10$\pm$0.01 in NLTE. Both LTE and NLTE parameters agree well with the photometric and asteroseismic parameters from \citet{Worley2020} within their error bars, however our NLTE parameter derivations agree better for both \teff\ and, particularly for \logg.

\subsection{CHARA Interferometry Stars}\label{sssec:3.2.4}
Finally, we apply and test \lotus\ on a sample of stars with  effective temperatures derived homogeneously from interferometric angular diameters observed by CHARA from \citet{Karo2020, Karo2021a, Karo2021b}. \citet{Karo2020} presented a study of interferometric observations of ten late-type metal-poor dwarfs and giants, whereas \citet{Karo2021a, Karo2021b} showed a similar analysis for several metal-rich stars. For three of their stars (namely, HD122563, HD140283, and HD175305), we have already presented their analysis in Section\,\ref{sssec:3.2.1} and \ref{sssec:3.2.2}. We, therefore, do not include the analysis of these stars again in this section. EW measurements of the total 12 stars were adopted from different literature sources including \citet{Takeda2005, Morel2014, Heiter2015, Takeda2015, Liu2020}.
%We also select stars with existing EW observation in previous studies. For dwarfs, they are HD131156 ($\xi$ Boo), HD146233 (18 Sco), HD186408 (16 Cyg A), HD186427 (16 Cyg B), HD190360, HD161797 and HD207978 (15 Peg); For giants and subgiants, they are: HD121370 ($\eta$ Boo), HD175955, and HD188512 ($\beta$ Aql).
The average uncertainties of the interferometric \teff\ estimated by \citet{Karo2020,Karo2021a, Karo2021b} are within 1$\%$. The authors derived their \logg\ values by fitting Dartmouth isochrones \citep{Dotter2008} to their interferometric \teff. They thus derive median uncertainties for \logg\ of 0.09 for their metal-poor star sample, 0.05 for their dwarf sample, and 0.07 for the giants/subgiants sample. 
%We derive the LTE and NLTE stellar parameters of the stars using \lotus. Comparisons of our results for \teff\ and \logg\ as compared to \citet{Karo2020,Karo2021a, Karo2021b} are shown in Figure \ref{fig:CHARA_v0}.

We find that on average the differences in \teff\ and \logg\ between the reference values \citep{Karo2020,Karo2021a, Karo2021b} and those derived with \lotus\ are within 26\,K and 0.01 in NLTE, respectively. The NLTE parameters derived for all the stars agree within 1$\sigma$ as compared to the reference values, except for the \logg\ value of HD121370 with [Fe/H]=0.29, which deviated by 0.12 for LTE and 0.18 for NLTE. \citet{Heiter2015}, however, reported a seismic \logg\ for this star of 3.83$\pm$0.02 dex, which agrees very with our NLTE result of 3.91$\pm$0.06 within 1$\sigma$.
%Below, we present a more detailed  discussion of our results, highlighting the obtained NLTE corrections for all parameters.    

\begin{figure*}[ht!]
\plotone{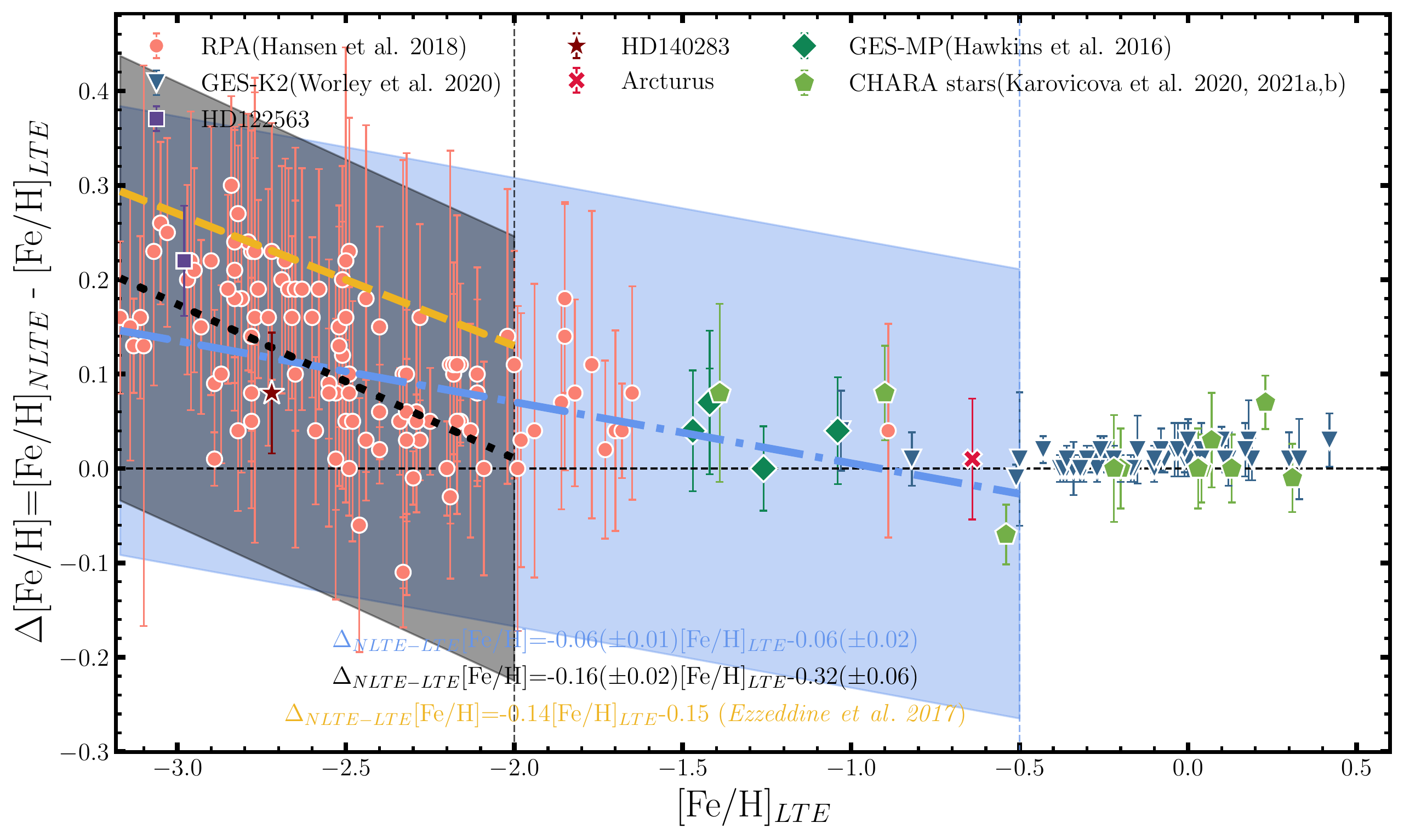}
\caption{NLTE corrections $\Delta$\feh\ versus \feh$_{\mathrm{LTE}}$ determined for this work's star sample using \lotus. Different markers indicate stars from different studies as indicated. The Yellow dashed line is a linear fit from \citet{Ezzeddine2017} for stars with [Fe/H]$<$-4.0. The Blue dot-dash line is our fit to the stars for \feh$<-0.5$, and the black dotted line is our fit to the stars with \feh$<-2.0$. \label{fig:nlte_corrections_feh}}
\end{figure*}

\section{NLTE corrections}\label{sec:application}
The stars presented in Section\,\ref{sec:test} for testing \lotus\ on benchmark stars with \teff\ and \logg\ derived using non-spectroscopic methods cover a limited range of metallicities (mostly with $\feh>-1.0$), as well as mainly dwarfs and sub-giant stars. However, NLTE stellar parameter ``corrections'' (also known as NLTE effects, defined as $\Delta$\,NLTE = parameter(NLTE) - parameter(LTE)) have been shown in the literature to be more significant and important for evolved (giants and supergiant) metal-poor stars \citet{Lind2012,Ezzeddine2017}. Thus, to be able to test our derived NLTE ``corrections'' on a larger, more metal-poor sample of giant stars, we derive LTE and NLTE stellar parameters for the $R$-process Alliance giant metal-poor star sample from \citet{Hansen2018}, who derived the stellar parameters of 107 metal-poor stars ($\feh<-1.0$) selected based on their $r$-process enhancement from several surveys. The EW measurements for \fei\ and \feii\ lines were adopted from \citet{Hansen2018}, measured from their du Pont spectroscopic observations. We thus add the RPA sample stars to our test sample stars presented in Section\,\ref{sec:test}, resulting in a significant sample covering a wide representative range of stellar parameters from \teff=4000 to 6500\,K, \logg=0.0 to 4.5, and \feh=$-3.0$ to $-0.5$. The derived stellar parameters for the RPA stars, in LTE and NLTE, are also listed in Table\,\ref{tab:derived_parameters}.

%In Figure\,\ref{fig:isochrones}, we show \logg\ versus \teff\ color-coded as a function of [Fe/H] for all the stars considered in Sections\ref{sec:test}, as well as the RPA sample on a Kiel diagram, for the parameters derived in LTE (left panel) and NLTE (right panel), respectively. Isochrones from MIST \citep{Choi2016} with a fixed age of 8\,Gyrs, [$\alpha$/Fe]\,=\,$+$0.4 and [Fe/H] ranging from $-3.0$ to $+$0.5 dex with steps of 0.5 dex are also overplot on the same figure. Compared with LTE values, NLTE values can be more coincident with isochrones of ages$=$8Gyr. While some K giant metal poor and  and F dwarf metal poor stars have larger \logg. 

We derive the NLTE corrections for [Fe/H] (denoted as $\Delta$\feh) for the full sample of stars. We plot the results as a function of $\feh_{\mathrm{LTE}}$ in Figure\,\ref{fig:nlte_corrections_feh}. As expected,  $\Delta$\feh\ increases toward lower metallicities. Such effects have been shown in multiple previous studies as well (for e.g., \citealp{M2011,Lind2012,Bergemann2012,Amarsi2016,Ezzeddine2017}). \citet{Ezzeddine2017} derived a linear relation between the NLTE correction for \feh, $\Delta$\feh, and \feh(LTE) from 20 ultra metal-poor stars with [Fe/H]$<-4.0$, such as:
\begin{equation}
    \Delta\mathrm{[Fe/H]} = -0.14\mathrm{[Fe/H]}_{\mathrm{LTE}}-0.15
\end{equation}

They found that their relationship can also be extended for metal-poor benchmark stars at $-4.0<$\,[Fe/H]\,$<-2.0$.  It is therefore useful to re-derive this equation using our full stellar sample analyzed uniformly using \lotus, for comparison. Following \citet{Ezzeddine2017}, we re-derive the relation between $\Delta$\feh\ and \feh(LTE) using our sample of stars . We re-derive the relation by (i) choosing only the stars with [Fe/H](LTE)$<-0.5$, and (ii) using the stars with [Fe/H](LTE)$<-2.0$. We thus derive $\Delta$\feh =    
\begin{equation}
\begin{cases}
         -0.06(\pm0.01)\mathrm{[Fe/H]}_{\mathrm{LTE}} - 0.06(\pm0.02),& \\ \mathrm{for~ [Fe/H]}_{\mathrm{LTE}}<-0.5 \\
         -0.16(\pm0.02)\mathrm{[Fe/H]}_{\mathrm{LTE}} -0.32(\pm0.06),& \\ \mathrm{for~ [Fe/H]}_{\mathrm{LTE}}<-2.0&
\end{cases}
\end{equation}

The two relations from these equations for $\Delta$\feh, as well as that derived in \citet{Ezzeddine2017}, are shown in Figure \ref{fig:nlte_corrections_feh}, fit to our test sample stars from Section\,\ref{sec:test} as well as the RPA sample.  We find that even though the three relations agree within uncertainties (colored bands in Figure\,\ref{fig:nlte_corrections_feh}), they can, however, yield slightly different NLTE corrections depending on the metallicity range for which they were fit. For e.g., our relation derived using stars with [Fe/H]$<-2.0$ underestimates the NLTE \feh\ correction as compared to \citet{Ezzeddine2017} by $\sim 0.1$\,dex at \feh\,=\,$-2.0$, while using the relation derived with [Fe/H]$<-0.5$ can underestimate the NLTE correction up to 0.2\,dex. This demonstrates that while these relations can provide useful first-order estimates of the NLTE corrections, a complete NLTE analysis (e.g., using \lotus) is needed for precise estimation of the corrections, as the former can be dependent on the incomplete sample of stars or metallicity range.         

\begin{figure*}[ht!]
\plotone{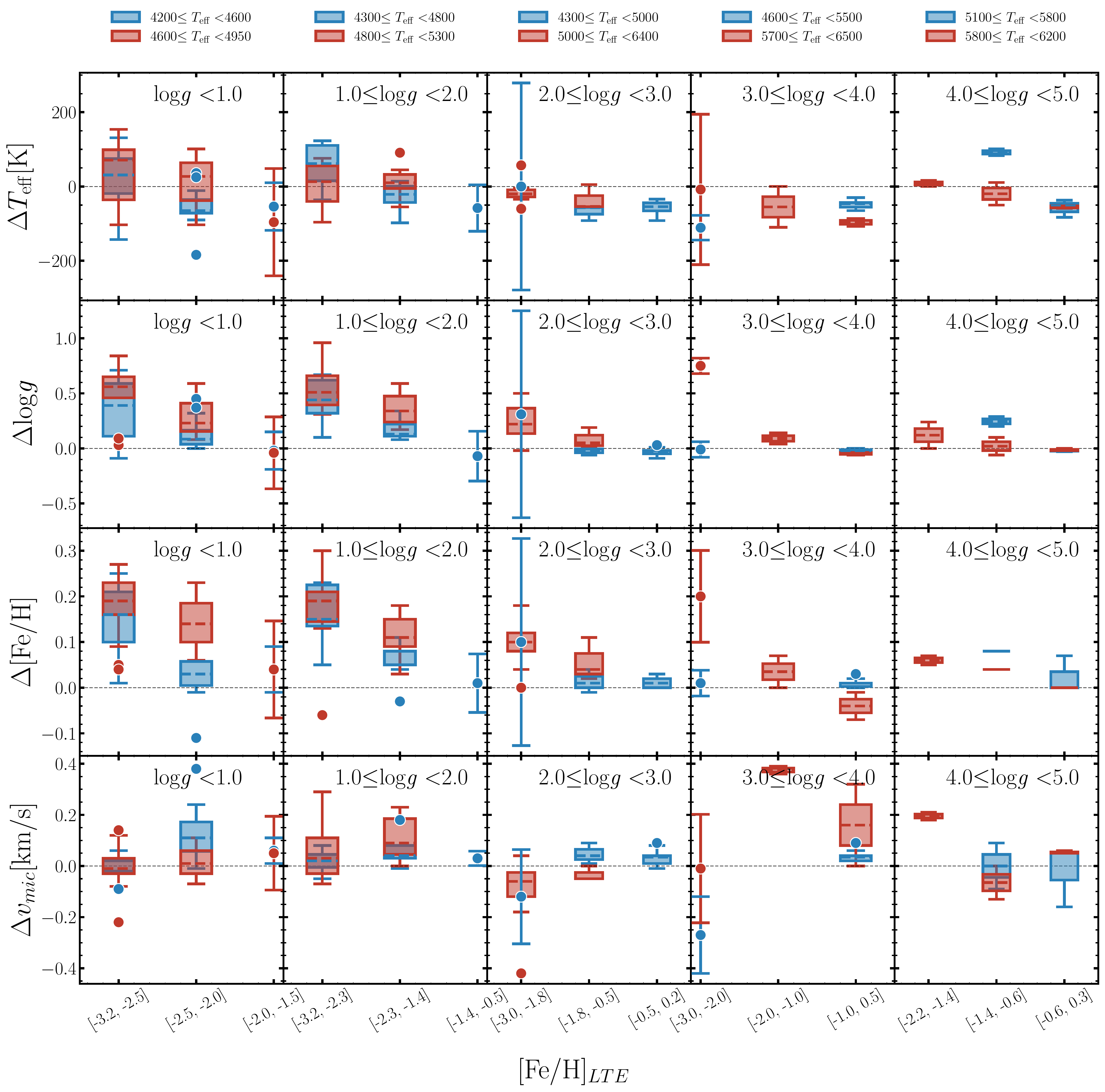}
\caption{Box plots showing the NLTE corrections of stars considered in this work for stellar parameter \teff, \logg, [Fe/H], and \vt\ as derived using \lotus. Box plots are color-coded for two ranges of \teff for each \logg range, indicated at the top of each column. For intervals with only one star from our sample, boxes are replaced with dots and their error bars. \label{fig:nlte_corrections_all}}
\end{figure*}

Similarly, we derive $\Delta$NLTE corrections for \teff, \logg, and \vt\ for our sample stars. Again, we define the NLTE correction for \teff\ as $\Delta$\teff\,=\,\teff(NLTE) - \teff(LTE), for \logg\ as $\Delta$\logg\,=\,\logg(NLTE) - \logg(LTE), and for \vt\ as $\Delta$\vt\,=\,\vt(NLTE) - \vt(LTE). To best represent these corrections as a function of the other stellar parameters, we divide our sample into box plots grouped into different \teff\ and \logg\ bins, as a function of \feh\ on the x-axis. The results are shown in Figure\,\ref{fig:nlte_corrections_all}.

In what follows we discuss the NLTE corrections we obtained for each parameter as a function of the other parameters, namely \teff, \logg\, and \feh. We also compare our result to the theoretical NLTE stellar parameter corrections derived by \citet{Lind2012}. We note, however, that our comparisons are affected by the fact that our analyzed stars do not cover exactly the same parameter space as theirs, and that our stellar parameters have been derived by \lotus\ simultaneously, i.e., taking into account their inter-dependencies, as explained in details in Section\,\ref{sec:method}, whereas \citet{Lind2012} derived their $\Delta$ NLTE corrections for each parameter independently, by fixing all the others.          
As shown in Figure\,\ref{fig:nlte_corrections_all}, we find that the \teff\ derived in NLTE are generally higher than those in LTE for very metal-poor ([Fe/H]$<-2.5$) super-giants and giant stars (\logg$<2.0$), with NLTE corrections up to 100\,K, whereas lower \teff\ are obtained in NLTE as compared to LTE as \logg\ and \feh\ increases.  \citet{Lind2012}, who derived their \teff\ using both excitation and ionization equilibrium by fixing gravity and other stellar parameters in the process. Similarly, they also restrict their \fei\ line transitions to certain cutoffs, choosing only lines with EP$>3.5$\,eV. They found that their ionization $\Delta$\teff\ yielded lower LTE \teff\ as compared to NLTE (see their Figure\,5), whereas their excitation $\Delta$\teff\ yielded negligible positive corrections for \teff$<4500$. More pronounced negative corrections were obtained for horizontal branches, supergiants, and extremely metal-poor stars, which are not covered in our sample stars. As noted above, our derived \teff\ using \lotus\ implement the contribution from both excitation and ionization equilibrium, by taking into account their \teff-\logg\ inter-dependencies.
In that context, a thorough comparison of our $\Delta$\teff\ results to either excitation or ionization theoretical corrections derived by \citet{Lind2012} are not very useful, however, similar to their results we generally find that the NLTE corrections for the bulk of our stars are affected by less than 50\,K (and up to 100\,K) in the considered parameter range, which is within our derived uncertainties. 

We also derive $\Delta$\logg\ for our sample stars. As expected, the NLTE corrections for \logg\ increase towards lower gravities, lower metallicities, and higher temperatures. On average, our corrections are $\sim +0.3-0.5$ for [Fe/H]$<-2.0$, $\sim +0.1-0.3$ for $-2.0<$[Fe/H]$<-1.0$, for giants and supergiant stars with \logg$<2.0$, with outliers reaching up to 1.0\,dex. For stars with \logg$>2.0$, the NLTE corrections can be up to $\sim$0.3 depending on metallicities.
These results broadly agree with the theoretical corrections derived in \citet{Lind2012}. It can thus be concluded that LTE analyses can strongly underestimate the surface gravities of stars, particularly for warmer metal-poor giants, and should thus be reliably derived in NLTE for reliable consequent stellar population analyses.

Finally, the NLTE corrections for \vt\ are generally small and are within $2.0$\,\kms\ from those derived in LTE. The corrections are mainly positive for lower metallicity giants and supergiant stars, as in particular their \fei\ lines are more strongly affected by NLTE, which leads to lower microturbulent velocities in LTE
as compared to NLTE. Our results are generally consistent with the $\Delta$\vt\ derived by \citet{Lind2012} from their theoretical models. 

\begin{figure*}[ht!]
\plotone{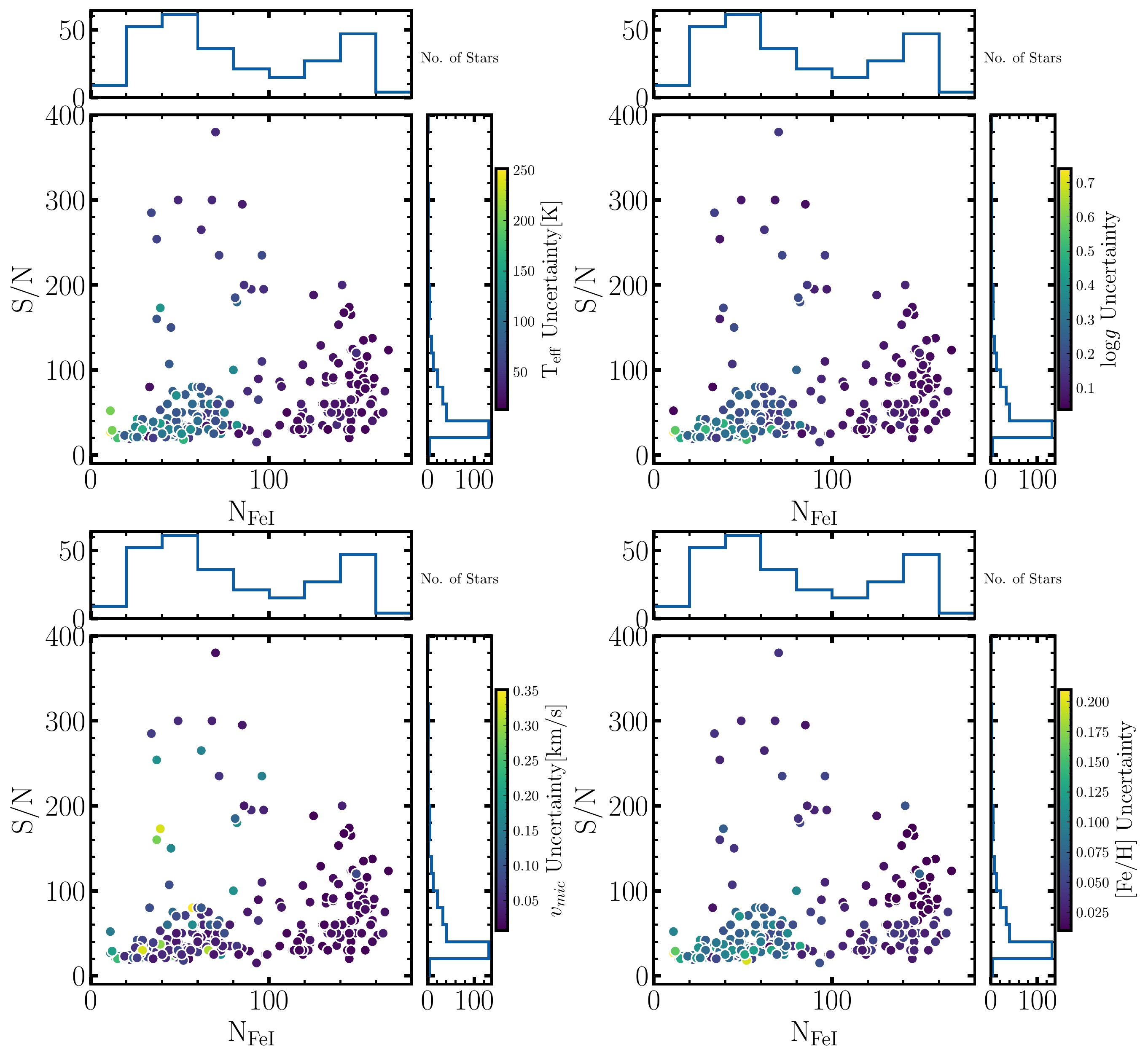}
\caption{Derived stellar parameter uncertainties as a functions of the number of Fe I lines and S/N ratios in our sample of stars from Section\,4, in addition to metal-poor stars with high S/N from \cite{Ezzeddine2020}.  %\sout{Uncertainties of the derived atmospheric stellar parameters of the stars considered in this work versus the number of iron lines color-coded for LTE (blue) and NLTE (red), respectively. Empty circles correspond to \fei\ lines and crosses for \feii\ lines. Close-up inset plots show the inner close region where most of the \feii\ lines are.}
\label{fig:precision}}
\end{figure*}

%\begin{figure*}[ht!]
%\plotone{accuracy.pdf}
%\caption{Differences between the atmospheric stellar parameters derived using \lotus\ of the stars considered in this work and the reference values in \citet{Hansen2018}, displayed versus the number of iron lines color-coded for \fei\ and \feii\ lines, in LTE (blue) and NLTE (red) respectively.\label{fig:accuracy}}
%\end{figure*}

\section{Caveats}\label{sec:caveats}
We have so far described and demonstrated that \lotus\ can be used as a reliable tool to derive NLTE and LTE atmospheric stellar parameters. There are, however, some caveats that warrant pointing out which can affect the results, particularly for stars occupying certain parameter spaces in our grids.

\paragraph{\it EW Interpolation} A key component of \lotus\ is its interpolation module that expresses the EW as a function of stellar parameters, using the generalized curve of growth (GCOG). As explained in Section\,\ref{sssec:select_interpolator}, we conduct thorough tests to choose the best polynomial interpolator for each spectral type. However, we find that some models are not adequate to reliably fit the GCOG, particularly for parameter regions where EW variations have smaller dependencies on stellar parameters. This is demonstrated in Figure\,\ref{fig2}, where the EW of the \fei\ line at 4077\,{\AA} varies less strongly with \vt\ at  higher \teff\ and lower \logg.
%For example, the \fei\ line at 3608.86\,{\AA} with EP=1.01\,eV, shows a difference between theoretic computed NLTE EW directly using \texttt{MULTI2.3} and \texttt{LOTUS} interpolated NLTE EW of 20\,m{\AA} at \teff\,$=$4000 and \feh\,$=-0.5$, while the same value can be as low as 5\,m{\AA} at the same \teff\ with \feh\,$=-3.5$.
Thus, the iron abundances derived from these interpolated models can deviate up to 0.5\,dex in abundances from those computed directly from \texttt{MULTI2.3} output (i.e, non interpolated EWs). Such large offsets are likely due to lower than optimal orders of polynomials chosen for some spectral types in these parameter regions (see Section\,\ref{sssec:select_interpolator} for details), where the EW of those lines not being sensitive to the variation of atmospheric stellar parameters, or being too small (close to zero) at most atmospheric stellar parameters. We emphasize though, that the number of such lines is limited and should not impact the statistical properties of the interpolation performance for the lines considered in our linelist, as shown in Figure \ref{fig:4}. 
%We do not drop these lines from our final linelist in Table\,\ref{tab:linelist} as the uncertainties can only affect a few \fei\ lines in a limited part of the parameter space, which are then automatically excluded as outliers in the final optimization during sigma clipping.
%In future \lotus\ updates, we plan to perform more detailed investigation into what parameter regions these lines occupy and how to improve interpolation of EW for these lines in the code. 

% These need to go under this paragraph. EW extraction will blend the target line into arbitrary several lines in the spectrum and depth of absorption will be shallow in lower spectral resolution \citep{Kang2012, BC2014}. 
%Except for the blend of lines, other factors like re-normalization of spectra and whether interpolation of atmospheric models or interpolation of simulated spectra/EW from original atmospheric models, can affect the performances of EW method. We recommend readers to \citet{BC2019} for details.

\paragraph{Number of Fe I lines and S/N  ratios} The derived atmospheric stellar parameters could be affected by the number of \fei\ and \feii\ lines utilized in the parameter optimization in \lotus. We, therefore, investigate the derived uncertainties of the stellar parameters using \lotus\ as a function of the number of Fe I lines and the S/N ratio as shown in Figure\,\ref{fig:precision}.
%\textbf{while we find the number of \feii\ lines has less impact - there are much less \feii\ lines in FGK spectra anyway}. 
To demonstrate our results over a representative statistical sample of stars covering a wide range of S/N ratios, we use the full star sample from Section\,\ref{sec:application}, supplemented by metal-poor stars from \cite{Ezzeddine2020} which cover higher S/N values (S/N$>$100) than those in \cite{Hansen2018}.
%from these two aspects statistically, we include stars within our parameter grid from \cite{Ezzeddine2020}, as the initial metal-poor sample from \citet{Hansen2018} did not have a sufficient sample of high S/N ratio.}
We can see in Figure \ref{fig:precision} that the uncertainties derived for each stellar parameter as a function of the number of \fei\ and S/N ratio decrease as a function of increasing the number of \fei\ and increasing the S/N ratio.
%\sout{\feii\ lines and are much more strongly affected by the number of \fei\ lines, $N_{\fei}$ in particular.} 
For stars with $N_{\mathrm{lines}}>100$ or stars with $N_{\mathrm{lines}}<100$ and S/N$>$100, the uncertainty are $<50$\,K for \teff,  $<0.1$\,dex for \logg, $<0.05$\,dex for [Fe/H] and $<0.02$\,\kms\ for \vt.
For stars with $N_{\mathrm{lines}}<100$ and S/N$<$60, however, the uncertainties can reach up to 200\,K for \teff,  0.5\,dex for \logg, 0.3\,dex for [Fe/H] and 0.2\,\kms\ for \vt. It is notable to mention, though, that higher uncertainties are more strongly correlated to lower S/N ($<40$). We thus recommend that users try to utilize "good-quality" EW measurements for their \fei\ (and \feii) lines to obtain reliable results and smaller uncertainty derivation in \lotus. This is particularly useful if the number of measured EW lines for a star is too small ($<100$). 
Additionally, as the number of \feii\ lines usually detected in cool FGK stellar spectra can be smaller than the number of \fei\ lines (on the order of 5-25 lines, depending on the \feh\ of the star), we moreover recommend that users try to maximize the number of \feii\ EW line measurements in the stars, if possible, to increase the precision of the parameter determination. 
In future work, we plan to test \lotus\ on a sample of low resolution spectroscopic observations (as compared to high-resolution data for the same stars) to accurately quantify the effects of the spectral resolution on the results of the stellar parameters derived by \lotus.

%We also investigate the differences between our derived parameters to reference values from \citet{Hansen2018} for the metal-poor sample as a function of $N_{\mathrm{lines}}$. In Figure\,\ref{fig:accuracy}, we notice that the uncertainties are mainly driven by the \fei\ lines. We also note that the dispersion is higher in LTE as compared to NLTE, where differences are on average within $200$\,K for $N_{\mathrm{lines}}<50$ and can even reach up to $200$\,K, whereas \logg dispersion is within 0.4 for $N_{\mathrm{lines}}<50$, and can reach up to 0.8\,dex. For NLTE, the average dispersion is within 150\,K for \teff\ and 0.2 for \logg. Similarly, for $N_{\mathrm{lines}}<50$ the dispersion can reach up to 0.15\,dex in \feh\ and 0.3\kms\ in \vt. The differences decrease to $\sim$ 50\,K for \teff,  0.2\,dex in \logg, 0.05\,dex in \feh\ and 0.05\,\kms\ in \vt for $N_{\mathrm{lines}}>50$. 

\paragraph{K-type stars} The atmospheric stellar parameters, particularly \logg, derived for K-type stars using \lotus\ can have larger uncertainties than those derived for other spectral types. \citet{Tsantaki2019} noted that \logg\ determined for K-type stars using ionization balance between \fei\ and \feii\ lines are underestimated depending on the choice of \feii\ lines used in the optimization. %They argued that the derived \logg\ can strongly depend on the choice of \feii\ lines and thus must be thoroughly selected.
We investigate this effect on our GES-K2 sample, in which most of the stars are K-type giants or sub-giants. We find that both our NLTE and LTE results can indeed overestimate the surface gravities up to 0.1 dex as compared to asteroseismic values, depending on the selection of \feii\ lines.

\paragraph{3D models}
Since our atmospheric stellar models are limited to 1D, our determined iron abundances might suffer from 3D effects due to atmospheric inhomogeneities,  horizontal radiation transfer, as well as the differences in the mean temperature stratification \citet{Amarsi2016} between 1D and 3D models. The latter can lead to underestimated abundances derived from \fei\ and \feii\ lines, on average up to $0.1$ dex in 1D, NLTE as compared with Fe abundances derived using 3D, NLTE. Such effects are, however, more strongly pronounced for \feii\ lines than \fei\ as explained in \citet{Amarsi2016}. We find that adding a $0.1$ dex correction for our 1D, NLTE \feh\ for HD122563 brings it closer to the 3D, NLTE \feh\ derived in \citep{Amarsi2016}. 

On the other hand, we find that a difference of $0.2$\,dex as compared to \citet{Amarsi2016} is still present when adding the same \feh\ correction for HD140283. This can be explained by the stellar parameter differences adopted in both studies, as well as the difference between linelists. However, as noted in \cite{Amarsi2016} and \citet{amarsi2022}, 3D corrections can nevertheless be smaller than the NLTE corrections determined for low metallicity evolved stars. Therefore, including 1D, NLTE corrections in the determination of \fei\ and \feii\ lines can significantly improve the derived stellar parameters as compared to 1D, LTE.

\section{Summary \& Conclusions}\label{sec:summary}

We present the open-source code, \lotus, developed to automatically derive fast and precise atmospheric stellar parameters (\teff, \logg, \feh\ and \vt) of stars in 1D, LTE, and 1D, NLTE using \fei\ and \feii\ equivalent width measurement from stellar spectra. 

\lotus\ implements a generalized curve of growth (GCOG) interpolation to derive abundances from a pre-computed grid of theoretical NLTE EW in a high-resolution and dense spectral parameter space. The GCOG takes into account the EW dependencies on stellar atmospheric parameters. A global differential evolution optimization algorithm, tailored to the spectral type of the star, is then used to derive the fundamental stellar parameters. In addition, \texttt{LOTUS} can be used to estimate precise uncertainties for each stellar parameter using a well-tested Markov Chain Monte Carlo (MCMC) algorithm.

We tested our code on several benchmark stars and stellar samples with reliable non-spectroscopic (from asteroseismic, photometric, and interferometric observations) measurements for a wide range of parameter space typical for FGK stars. We find that our NLTE-derived stellar parameters for \teff\ and \logg\ are within 30\,K and 0.1 dex for benchmark stars including the metal-poor standard stars, HD140283 and HD122563, as well as Arcturus. We also test \lotus\ on a large sample of $Gaia$-ESO (GES) dwarf stars from \citet{Hawkins2016} and GES-K2 with asteroseismic gravities from K2, as well as stars with CHARA interferometric observations. Similarly, we find that \lotus\ performs very well in reproducing the non-spectroscopic \teff\ and \logg\ on average within $<20$\,K and $<0.1$\,dex in NLTE, as compared to $\sim 70$\,K and $\sim 0.2$\,dex in LTE.

Moreover, we apply \lotus\ on a large sample of metal-poor stars from the R-Process Alliance (RPA, \citealt{Hansen2018}). We use the RPA sample as well as the test sample stars to derive NLTE stellar parameter corrections for our stars and review the $\Delta$\feh\ versus \feh(LTE) relation derived in \citet{Ezzeddine2017}. We find that our NLTE corrections agree with theoretical corrections predicted by \citet{Lind2012}, where general trends of corrections were found to increasing toward metal-poor evolved stars, as expected.

We test \lotus\ thoroughly for its performance for different spectral types and parameter space. We find that despite some caveats discussed in Section\,\ref{sec:caveats}, \lotus\ can overall be reliably used to provide fast and accurate NLTE derivation for a wide range of stellar parameters, especially metal-poor giants which can be strongly affected by deviations from LTE. We thus strongly recommend that the community use it to apply for the spectroscopic analyses of stars and stellar populations.

We provide open community access to \lotus, as well as the pre-computed interpolated LTE and NLTE EW grids available on Github \href{https://github.com/Li-Yangyang/LOTUS}{\faIcon{github}}, with documentation and working examples on Readthedocs \href{https://lotus-nlte.readthedocs.io/en/latest/}{\faIcon{book}}.

\acknowledgments
We thank the anonymous referee for their useful comments that heloped improved the manuscript. R.E. acknowledges support from JINA-CEE (Joint Institute for Nuclear Astrophysics - Center for the Evolution of the Elements), funded by the NSF under Grant No. PHY-1430152. Y.L. and R.E. acknowledge support from NSF grant AST-2206263.
The authors acknowledge the University of Florida Research Computing for providing computational resources and support that have contributed to the research results reported in this publication. URL: \href{http://researchcomputing.ufl.edu}{UFRC}. 

\software{astropy \citep{Astropy2013, Astropy2018},
          numpy \citep{Numpy2011, Numpy2020},
          pandas \citep{pandas},
          scikit-learn \citep{sklearn},
          scipy \citep{scipy},
          sympy \citep{sympy},
          matplotlib \citep{matplotlib},
          h5py \citep{hdf5},
          PyMC3 \citep{pymc3},
          theano \citep{theano},
          corner \citep{corner},
          numdifftools \citep{numdifftools}
          }


\clearpage
\appendix

\section{Linelist of \fei\ and \feii\ in \lotus}

\startlongtable
\begin{deluxetable}{c c c c}
\twocolumngrid
\tablewidth{0pt}
\tabletypesize{\footnotesize}
\tablecaption{\fei\ and \feii\ linelist selected for \texttt{LOTUS} (see Section\, \ref{sec:linelist} for details) }
\tablehead{\colhead{Species} & \colhead{$\lambda$($\AA$)}  & \colhead{E$_{low}$(ev)} & loggf \label{tab:linelist}}
\startdata
\fei    & 3440.61 & 0.00  & $-$0.67  \\
\fei    & 3440.99 & 0.05 & $-$0.96  \\
\fei    & 3447.28 & 2.20  & $-$1.02  \\
\fei    & 3450.33 & 2.22 & $-$0.90   \\
\fei    & 3451.91 & 2.22 & $-$1.00   \\
\fei    & 3476.70  & 0.12 & $-$1.51  \\
\fei    & 3490.57 & 0.05 & $-$1.10   \\
\fei    & 3497.84 & 0.11 & $-$1.55  \\
\enddata
\end{deluxetable}

\section{Derived Stellar Parameters of the Target Stars in This Work}

\begin{longrotatetable}
\begin{deluxetable}{l c c c c c c c c c c c c c c c c}
\tablewidth{700pt}
\tabletypesize{\scriptsize}
\movetabledown=25mm
\tablecaption{\label{tab:stell_param} Derived Stellar Atmospheric Parameters of the Target Stars\label{tab:derived_parameters}}
\tablehead{
\\
& \multicolumn{6}{c}{LTE}   & & 
\multicolumn{6}{c}{NLTE} \\
\cline{2-8} \cline{10-16} 
\colhead{Star Name} & \colhead{\teff} & \colhead{\logg} & \colhead{[Fe/H]} & \colhead{\vt} & \colhead{N$_\mathrm{Fe_I}$} & \colhead{N$_\mathrm{\fei}$} & \colhead{$\chi_{\mathrm{Fe_I}}$cutoff} & & \colhead{\teff} & \colhead{\logg} & \colhead{[Fe/H]}  & \colhead{\vt}  & \colhead{N$_\mathrm{\fei}$} & \colhead{N$_\mathrm{\feii}$} & \colhead{$\chi_{\mathrm{\fei}}$cutoff}\\
 & \colhead{(K)} & \colhead{(cgs)} & \colhead{(dex)} & \colhead{(km s$^{-1}$)} & & & \colhead{(ev)} & & \colhead{(K)} & \colhead{(cgs)} & \colhead{(dex)}  & \colhead{(km s$^{-1}$)} & & & \colhead{(ev)} &
}
\startdata
HD122563          & 4528$\pm$42  & 0.66$\pm$0.12 & $-$2.98$\pm$0.03 & 1.84$\pm$0.03 & 72          & 20      & 2.7 &  & 4620$\pm$59  & 1.21$\pm$0.19 & $-$2.76$\pm$0.05 & 1.89$\pm$0.06 & 72          & 20      & 2.7 \\
HD140283          & 5689$\pm$59  & 3.29$\pm$0.17 & $-$2.72$\pm$0.04 & 2.09$\pm$0.14 & 96          & 15      & 2.0 &  & 5681$\pm$67  & 3.58$\pm$0.16 & $-$2.64$\pm$0.05 & 2.17$\pm$0.16 & 96          & 15      & 2.0 \\
Arcturus          & 4364$\pm$47  & 1.68$\pm$0.17 & $-$0.64$\pm$0.05 & 1.76$\pm$0.02 & 70          & 8       & 0.0 &  & 4306$\pm$41  & 1.61$\pm$0.15 & $-$0.63$\pm$0.04 & 1.79$\pm$0.02 & 70          & 8       & 0.0 \\
\hline
\multicolumn{16}{c}{GES Metal-Poor Stars \citep{Hawkins2016}} \\
\hline
HD102200          & 6167$\pm$52  & 3.98$\pm$0.09 & $-$1.26$\pm$0.02 & 0.62$\pm$0.09 & 33          & 4       & 2.5 &  & 6057$\pm$75  & 4.02$\pm$0.09 & $-$1.26$\pm$0.04 & 0.98$\pm$0.28 & 37          & 4       & 2.5 \\
HD106038          & 6108$\pm$51  & 4.05$\pm$0.07 & $-$1.42$\pm$0.02 & 0.59$\pm$0.07 & 35          & 5       & 2.5 &  & 6124$\pm$60  & 4.29$\pm$0.08 & $-$1.35$\pm$0.03 & 0.8$\pm$0.18  & 37          & 5       & 2.5 \\
HD175305          & 5113$\pm$84  & 2.51$\pm$0.25 & $-$1.47$\pm$0.05 & 1.61$\pm$0.2  & 45          & 6       & 2.5 &  & 5059$\pm$70  & 2.56$\pm$0.2  & $-$1.43$\pm$0.04 & 1.61$\pm$0.17 & 45          & 6       & 2.5 \\
HD201891          & 6036$\pm$86  & 4.23$\pm$0.15 & $-$1.04$\pm$0.04 & 0.63$\pm$0.09 & 44          & 5       & 2.5 &  & 6047$\pm$81  & 4.33$\pm$0.14 & $-$1.0$\pm$0.04  & 0.63$\pm$0.09 & 44          & 5       & 2.5 \\
\enddata
\tablecomments{\tablenotetext{\dagger}{For HD121370, parameters are obtained with the average of runs inputting 3 different EW line lists: EPINARBO in \citet{Heiter2015}, BOLOGNA in \citet{Heiter2015} and \citet{Takeda2005}. In LTE, there are N$_{\mathrm{Fe_I}}$=79, 121, 46, N$_{\mathrm{Fe_II}}$=6, 9, 4 lines in these three line list respectively; In NLTE, N$_{\mathrm{Fe_I}}$=79, 121, 47, N$_{\mathrm{Fe_II}}$=6, 9, 4}.\\
(This table is available in machine$-$readable form.)}
\end{deluxetable}
\end{longrotatetable}

\bibliography{sample63}{}
\bibliographystyle{aasjournal}



\end{document}